
\input amstex
\documentstyle {amsppt}
\document
\nologo
\NoBlackBoxes
\magnification=1200
\centerline{\bf GROMOV--WITTEN 	CLASSES, QUANTUM COHOMOLOGY,}
\centerline{\bf AND ENUMERATIVE GEOMETRY}

\medskip

\centerline{M. Kontsevich, Yu. Manin}

\medskip

{\it Max--Planck--Institut f\"ur Mathematik, Gottfried--Claren--Str. 26,
53225, Bonn, Germany}

\vskip 1cm

\centerline{ABSTRACT}

\medskip

The paper is devoted to the mathematical aspects of topological
quantum field theory and its applications to enumerative problems
of algebraic geometry. In particular, it contains an axiomatic
treatment of Gromov--Witten classes, and a discussion of
their properties for Fano varieties. Cohomological
Field Theories are defined, and it is proved that
tree level theories are determined by their correlation
functions. Application to counting rational curves
on del Pezzo surfaces and projective spaces are given.

\vskip 1cm

\centerline{\bf \S 1. Introduction}

\bigskip

Let $V$ be a projective algebraic manifold.

\smallskip

Methods of quantum field theory recently led to a prediction of some
numerical characteristics of the space of algebraic curves in $V$,
especially of
genus zero, eventually endowed with a parametrization
and marked points. It turned out that an appropriate generating
function $\Phi$ whose coefficients are these numbers has a physical
meaning (``potential'', or ``free energy''), and its analytical
properties can be guessed with such a precision that it becomes
uniquely defined. In particular, when $V$ is a Calabi--Yau manifold,
$\Phi$ conjecturally describes a variation of Hodge structure of
the mirror dual manifold in special coordinates (see contributions
in [Y], [Ko], [Ma2]) which identifies $\Phi$ as a specific combination of
hypergeometric functions.

\smallskip

In this paper, we use a different tool, the so called ``associativity''
relations, or WDVV--equations (see [W], [D]),
in order to show that for Fano
manifolds these equations tend to be so strong that they can define
$\Phi$ uniquely up to a choice of a finite number of constants.
(For Calabi--Yau varieties these equations hold as well, but they
do not constraint $\Phi$ to such extent).

\smallskip

Mathematically, this formalism is based upon the theory of
{\it the Gromov--Witten classes}. In our setup, they form a collection
of linear maps $I^V_{g,n,\beta}:\ H^*(V,\bold{Q})^{\otimes n}\to
H^*(\overline{M}_{g,n},\bold{Q})$ that ought to be defined for all
integers $g\ge 0, n+2g-3\ge 0,$ and homology classes
$\beta \in H_2(V,\bold{Z})$ and are expected to satisfy a series
of formal properties as well as geometric ones. (Here $\overline{M}_{g,n}$
is the coarse moduli space of stable curves of genus $g$ with $n$
marked points).

\smallskip

In \S 2 of this paper, we compile a list of these formal properties,
or ``axioms'' (see subsections 2.2.0--2.2.8), and explain
the geometric intuition behind them (2.3.0--2.3.8). This is an
elaboration of Witten's treatment [W].

\smallskip

Unfortunately, the geometric construction of these classes to our knowledge
has not been given even for $V=\bold{P}^1.$ The most advanced
results were obtained for $g=0$ by the techniques of symplectic
geometry going back to M. Gromov (see [R], [RT]), but they fall
short of the complete picture. In 2.4 we sketch an algebro--geometric
approach to this problem based upon a new notion of {\it stable map}
due to one of us (M. K.)

\smallskip

The axiomatic treatment of \S 2 in principle opens a way to prove
this existence formally, at least for some Fano varieties $V$
and $g=0$. This is the content of $\S 3$ and the Reconstruction
Theorem 3.1, which basically says that Gromov--Witten classes
in certain situations can be recursively calculated. However
the equations determining these classes form a grossly overdetermined
family, so that checking compatibility at each step presents
considerable algebraic difficulties. The Second Reconstruction Theorem
8.8 shows that it suffices to check this compatibility
for codimension zero classes. This allows one to extend the construction
of [RT] from codimension zero to all tree level GW--classes.

\smallskip

The subject matter of \S 4 is the beautiful geometric picture encoded
in the potential function $\Phi$ constructed with the help of
zero--codimensional Gromov--Witten classes of genus zero.
Namely, over a convergence subdomain $M\subset H^*(V)$ (the cohomology
space being considered as a linear supermanifold) $\Phi$ induces
the following structures:

\smallskip

a). A structure of the (super)commutative associative algebra
with identity on the (fibers of the) tangent bundle $\Cal{T}H^*(V)$
depending on the point $\gamma \in H^*(V).$ The fibers
$\Cal{T}_{\gamma}H^*(V)$ were called by Vafa ``quantum cohomology
rings'' of $V$.

\smallskip

b). A flat connection on $\Cal{T}H^*(V)$ which was used by
B. Dubrovin [D] in order to show that the associativity
equations constitute a completely integrable system.

\smallskip

c). An extended connection on $\Cal{T}H^*(V)$ lifted to
$H^*(V)\times \bold{P}^1$ and its partial Fourier transform
which may define a variation of Hodge structure.

\smallskip

We show that the axioms for the Gromov--Witten classes
imply all the properties of $\Phi$ postulated in [D].

\smallskip

Together, \S 2 and \S 4 can be considered as a pedagogical
attempt to present the formalism of correlation functions
of topological sigma--models in a form acceptable for mathematicians
with algebro--geometric background.

\smallskip

A more ambitious goal of our treatment is to define a framework
for the conjectural interpretation of $H^*(V)$ as an
extended moduli space (see [Ko] and Witten's contribution
to [Y]).

\smallskip

In \S 5 we discuss examples.
Since from the enumerative geometry viewpoint the logic
of this discussion is somewhat convoluted, we try to describe it
here.

\smallskip

{\it Assuming} the existence of the relevant Gromov--Witten
classes we calculate the potential $\Phi$ and give the
recursive formulas for its coefficients whenever feasible.
{\it Assuming} in addition that these classes can be
constructed and/or interpreted along the lines of \S 2,
we state the geometric meaning of these numbers.

\smallskip

On the other hand, the potential $\Phi$ can be directly {\it defined}
by using the (numerical version of the) Reconstruction Theorem.
Then the redundancy of the associated equations translates
into a family of strange number--theoretical identities.
In principle, they can be also checked directly, without recourse
to the geometric context in which they arose. Until this is
done, they remain conjectural. We discuss del Pezzo surfaces
from this angle (cf. also [I]).

\smallskip

The last three sections are devoted to a description of a less
constrained structure of Cohomological Field Theory.
Roughly speaking, we forget about the dependence of our theory
on the target manifold $V,$ and retain only its part dealing
with moduli spaces.
In \S 6, we give two definitions of a CohFT and prove their equivalence.
One is modelled upon the axiomatics of Gromov--Witten classes,
another is based upon (a version of) operads.

\smallskip

This formalism is used in \S 7 for a description of the cohomology
of moduli spaces of genus zero. Keel in [Ke] described its ring
structure in terms of generators, the classes of boundary divisors,
and relations between them. We need more detailed understanding
of linear relations between homology classes of boundary strata
of any codimension, and derive from Keel's result a complete
system of such relations. (E. Getzler informed us that he
and R. Dijkgraaf obtained similar results).

\smallskip

Finally, in \S 8 we prove the second Reconstruction Theorem,
which allows us, in particular, to classify Cohomology
Field Theories via solutions of WDVV-equations, and to formally
prove the existence of GW--classes e.g., for projective spaces.
This theorem can be viewed as an instance of a general principle
that a quantum field theory can be completely recovered
from the collection of its Green functions.

\smallskip

The authors are grateful to the Max--Planck--Institut f\"ur
Mathematik in Bonn for the stimulating atmosphere in which
this work was done.

\newpage

\centerline{\bf \S 2. Gromov--Witten classes}

\bigskip

{\bf 2.1. Setup.} Let $V$ be a projective algebraic
manifold over $\bold{C}$ with canonical class $K_V.$

\smallskip

Denote by $B\subset H_2(V,\bold{Z})$ the semigroup consisting
of homology classes $\beta$ such $(L.\beta)\ge 0$ for all K\"ahler
$L$.

\smallskip

In what follows, we will often consider cohomology classes as
represented by differential forms, and then write e.g.
$\int_C c_1(T(V))$ instead of $(-K_V.C).$ Cup product is denoted
$\wedge .$

\medskip

{\bf 2.2. Definition.} A {\it system} (resp. {\it tree level system})
{\it of Gromov--Witten (GW) classes} for $V$ is a family of linear
maps
 $$
I^V_{g,n,\beta}:\ H^*(V,\bold{Q})^{\otimes n}\to H^*(\overline{M}_{g,n},
\bold{Q}) \eqno{(2.1)}
 $$
defined for all $g\ge 0, n\ge 0, n+2g-3\ge 0$ (resp. $g=0, n\ge 3$)
and satisfying the following axioms.

\smallskip

{\bf 2.2.0. Effectivity.} $I^V_{g,n,\beta}=0$ for $\beta \notin B.$

\smallskip

{\bf 2.2.1. $S_n$--covariance.} The symmetric group $S_n$ acts upon
$H^*(V,\bold{Q})^{\otimes n}$ (considered as superspace via
$\bold{Z}$ mod 2 grading) and upon $\overline{M}_{g,n}$
via renumbering of marked points. The maps $I^V_{g,n,\beta}$
must be compatible with this action.

\smallskip

{\bf 2.2.2. Grading.} For $\gamma \in H^i,$ put $|\gamma|=i.$
The map $I^V_{g,n,\beta}$ must be homogeneous of degree
$2(K_V.\beta )+(2g-2)\roman{dim}_{\bold{C}}V,$ that is
 $$
|I^V_{g,n,\beta}(\gamma_1\otimes \cdots \otimes \gamma_n)|=
\sum^n_{i=1}|\gamma_i|+2(K_V.\beta)+(2g-2)\roman{dim}_{\bold{C}}V.
\eqno{(2.2)}
 $$

Before stating the remaining axioms, let us introduce the following
terminology. Call a GW-class {\it basic} if it corresponds to the
least admissible values of $(n,\beta)$ that is, belongs to the
following list:
 $$
I^V_{0,3,\beta}(\gamma_1 \otimes \gamma_2 \otimes \gamma_3 );\
I^V_{1,1,\beta}(\gamma );\ I^V_{g,0,\beta}(1),\ g\ge 2, \eqno{(2.3)}
 $$
where $1$ is the canonical generator of $H^*(V,\bold{Q})^{\otimes 0}=
\bold{Q}.$ Call a class {\it new} if it is not basic, and if among
its (homogeneous) arguments $\gamma_i$ there are none with
$|\gamma |=0$ or $2.$ Finally, call the number
 $$
2(3g-3+n)-|I^V_{g,n,\beta}(\gamma_1\otimes \dots \otimes\gamma_n)|
 $$
the {\it codimension} of the class (recall that $\roman{dim}_{\bold{C}}
\overline{M}_{g,n}=3g-3+n).$ The classes of codimension zero are
especially important and are expected to express the number of solutions
of some counting problems ( see 2.3 below). Instead of such a class
$I^V_{g,n,\beta}$ we will often consider the corresponding number
$\langle I^V_{g,n,\beta}\rangle$ defined by
 $$
\langle I^V_{g,n,\beta}\rangle (\gamma_1\otimes \dots \otimes \gamma_n)=
\int_{\overline{M}_{g,n}}I^V_{g,n,\beta}(\gamma_1\otimes \dots
\otimes \gamma_n). \eqno{(2.4)}
 $$

Notice the following facts:

a). For $g=0,$ all basic classes have codimension zero, because
$\roman{dim}_{\bold{C}}\overline{M}_{0,3}=0.$

b). For a non-vanishing class, we have from (2.2):
 $$
(-K_V.\beta )-n-(g-1)\roman{dim}_{\bold{C}}V\le
\frac{1}{2}\sum_{i=1}^{n}(|\gamma_i|-2) \le
 $$
 $$
(-K_V.\beta)-(g-1)(\roman{dim}_{\bold{C}}V-3), \eqno{(2.5)}
 $$
with the second equality sign for codimension zero classes.
If the class is new, and $H^1(V)=0,$
the middle term of (2.5) must be non-negative,
and even $\ge n$, if $H^3(V)=0.$ Hence for $K_V=0, \roman{dim}_{\bold{C}}V=3,$
(e.g., Calabi--Yau threefolds) there are no non-vanishing new classes.
For Fano varieties, this inequality bounds $n$ if $\beta$ is fixed,
and $\beta$ if $n$ is fixed.

\smallskip

The next two axioms partially reduce the calculation of GW-classes
to that of basic ones and new ones.

\bigskip

{\bf 2.2.3. Fundamental class.} Let $e_V^0\in H^0(V)$ be the
identity in the cohomology ring (i.e. its dual homology class
is the fundamental class $[V]$). If the following class is not
basic, we have
 $$
I^V_{g,n,\beta}(\gamma_1\otimes \dots \otimes \gamma_{n-1}\otimes e^0_V)=
\pi^*_{n}I^V_{g,n-1,\beta}(\gamma_1\otimes \dots \otimes \gamma_{n-1})
\eqno{(2.6)}
 $$
where $\pi_n:\ \overline{M}_{g,n}\to \overline{M}_{g,n-1}$ is the map
forgetting the $n$--th section. In particular, (2.6) cannot be
of codimension zero unless it vanishes.

\smallskip

In addition, for basic classes with argument $e^0_V$ we have
 $$
\langle I^V_{0,3,\beta}\rangle (\gamma_1\otimes \gamma_2 \otimes e^0_V)=
\cases 0,\ \roman{if}\ \beta \ne 0;\\
\int_V\gamma_1\wedge \gamma_2,\ \roman{if}\  \beta = 0.
 \endcases \eqno(2.7)
 $$

\medskip

{\bf 2.2.4. Divisor.} If $I^V_{g,n,\beta}(\gamma_1\otimes \dots
\otimes \gamma_n)$ is a non-basic codimension zero class with
the last argument $\gamma_n,\ |\gamma_n|=2,$ then
 $$
\langle I^V_{g,n,\beta}\rangle (\gamma_1\otimes \dots \otimes
\gamma_n)=\int_{\beta}\gamma_n \cdot \langle I^V_{g,n-1,\beta}\rangle
(\gamma_1\otimes \dots \otimes \gamma_{n-1}).
 $$

More generally, for non-necessarily codimension zero classes,
 $$
\pi_{n*}(I^V_{g,n,\beta}(\gamma_1\otimes \dots \otimes \gamma_n))=
\int_{\beta}\gamma_n \cdot I^V_{g,n-1,\beta}(\gamma_1 \otimes
\dots \otimes \gamma_{n-1}).
 $$

\medskip

{\bf 2.2.5. Mapping to point.} By this catchword we describe the
situation when $\beta =0.$

\smallskip

For $g=0,$ the situation is simple: the only non-vanishing classes must
have $\sum_{i=1}^n |\gamma_i|=2\roman{dim}_{\bold{C}}V,$ and
 $$
I^V_{0,n,0}(\gamma_1 \otimes \dots \otimes \gamma_n)=
(\int_V \gamma_1\wedge \dots \wedge \gamma_n)e^0_{\overline{M}_{0,n}},
\eqno{(2.8)}
 $$
where $e^0_{\overline{M}_{0,n}}$ is the identity in
$H^*(\overline{M}_{0,n}).$

\smallskip

For $g=1,$ the non-vanishing classes have $\sum_{i=1}^n|\gamma_i|=0$ or $2$
and therefore are lifts of the following two basic ones:
 $$
I^V_{1,1,0}(e^0_V)=\chi (V)e^0_{\overline{M}_{1,1}}, \eqno(2.9)
 $$
 $$
|\gamma |=2:\ I^V_{1,1,0}(\gamma )=(\int_{V}\Cal{X}(V)\wedge\gamma )
e^2_{\overline{M}_{1,1}}, \eqno (2.10)
 $$
where $\chi (V)=\sum_i(-1)^ih^{i,0}(V),\ e^2_{\overline{M}_{1,1}}=
c_1(\Cal{O}(1))$ (recall that $\overline{M}_{1,1}\simeq
\bold{P}^1)$, and $\Cal{X}(V)$ is a certain characteristic
class of the tangent bundle of $V$ whose explicit description
we postpone to 2.4.2.

\smallskip

Finally if $g\ge 2$, then $I^V_{g,n,0}=0$ unless $\roman{dim}_{\bold{C}}V\le
3,$
and non-vanishing classes again can be described explicitly.

\medskip

{\bf 2.2.6. Splitting.} Fix $g_1,g_2$ and $n_1,n_2$ such that
$g=g_1+g_2,\ n=n_1+n_2,\ n_i+2g_i-2\ge 0.$ Fix also two
complementary subsets $S={S_1,S_2}$ of $\{1,\dots ,n\},$
$|S_i|=n_i.$ Denote by $\varphi_S:\
\overline{M}_{g_1,n_1+1}\times \overline{M}_{g_2,n_2+1}\to
\overline{M}_{g,n}$
the canonical map which assigns to marked curves
$(C_i;x^{(i)}_1,\dots ,x^{(i)}_{n_i+1}), i=1,2,$ their union
$C_1\cup C_2,$ with $x^{(1)}_{n_1+1}$ identified to $x^{(2)}_1.$
The remaining points are then renumbered by $\{1,\dots ,n\}$
in such a way that their relative order is kept intact, and
points on $C_i$ are numbered by $S_i.$

\smallskip

Finally, choose a homogeneous basis $\{ \Delta_a|a=1,\dots ,D\}$
of $H^*(V,\bold{Q})$ and put $g_{ab}=\int_V\Delta_a\wedge\Delta_b,
(g^{ab})=(g_{ab})^{-1}.$

\smallskip

The Splitting Axiom now reads:
 $$
\varphi^*_S(I^V_{g,n,\beta}(\gamma_1\otimes \dots \otimes \gamma_n))=
 $$
 $$
\varepsilon(S)\sum_{\beta =\beta_1+\beta_2}\sum_{a,b}
I^V_{g_1,n_1+1,\beta_1}((\otimes_{j\in S_1}\gamma_j)\otimes \Delta_a)
g^{ab}\otimes I^V_{g_2,n_2+1,\beta_2}(\Delta_b\otimes
(\otimes_{j\in S_2}\gamma_j)),
\eqno(2.11)
 $$
where $\varepsilon(S)$ is the sign of the permutation induced by $S$
on $\{\gamma_j\}$ with odd $|\gamma |.$

\smallskip

Notice that $\sum_{a,b}\Delta_a\otimes \Delta_bg^{ab}$ is the class of
the diagonal in $H^*(V\times V).$

\smallskip

The sum in (2.11) is finite because of the Effectivity Axiom 2.2.0.

\medskip

{\bf 2.2.7. Genus reduction.} Denote by $\psi :\ \overline{M}_{g-1,n+2}
\to \overline{M}_{g,n}$ the map corresponding to glueing together
the last two marked points. Then
 $$
\psi^*(I^V_{g,n,\beta}(\gamma_1\otimes \dots \otimes\gamma_n))=
\sum_{a,b}I^V_{g-1,n+2,\beta}(\gamma_1\otimes \dots \otimes \gamma_n
\otimes \Delta_a \otimes \Delta_b)g^{ab}.
\eqno(2.12)
 $$

\medskip

{\bf 2.2.8. Motivic Axiom.} The maps $I^V_{g,n,\beta}$ are induced
by certain correspondences in the Chow rings:
 $$
C^V_{g,n,\beta}\in C^*(V^n\times{\overline{M}}_{g,n}). \eqno(2.13)
 $$
We recall that $c\in C^*(V\times W)$ induces the map
 $$
H^*(V)\to H^*(W):\ \gamma \mapsto \pi_{W*}(\pi^*_V(\gamma )
\wedge [c]), \eqno(2.14)
 $$
where $\pi_V, \pi_W$ are projections of $V\times W$ to
$V,W$ and $[c]$ is the cohomology class of $c.$

\medskip

{\bf 2.3. Comments to the Definition.} When Gromov--Witten classes
are considered in the literature in an algebraic context,
it is usually assumed that $V$ is either Fano ($-K_V$ ample),
or Calabi-Yau ($K_V=0$) so that $(-K_V.\beta )\ge 0$ is automatically
satisfied for all algebraic homology classes. However, nothing
in the formalism forces us to postulate it. Notice
that for manifolds of general type (2.5) implies vanishing
of all $I^V_{g,n,\beta}$ with large enough $\beta .$

\smallskip

The simplest example of a tree level system of GW--classes
on $V$ is:
 $$
I^V_{0,n,\beta}(\gamma_1\dots\gamma_n)=
\cases 0, &\roman{if\ } \beta\ne 0,\\
(\int_V\gamma_1\wedge\dots\wedge\gamma_n)e^0_{\overline{M}_{0,n}}&
\roman{otherwise.}
\endcases
\eqno (2.15)
 $$
In addition, any system of GW--classes posesses an obvious
scaling transformation (if we allow to extend coefficients $\bold{Q}$
to $\bold{C}$):
$I^V_{0,n,\beta}\mapsto e(\beta )I^V_{0,n,\beta}$
where $e:\ B\to \bold{C}^*$ is a semigroup homomorphism.
If we put $e(\beta )=\roman{exp}(-t(\omega .\beta))$ for a K\"ahler
class $\omega ,$ then the scaling of any initial tree level
system tends to (2.15) as $t\to\infty .$ In terms of quantum
cohomology (see 4.5 below) (2.15) gives rise to the classical
cohomology ring, whereas $I^V_{0,n,\beta}$ supply the
instanton corrections.

\smallskip

Intuitively, one can imagine the geometry behind these corrections
$I^V_{g,n,\beta}$ as follows.
For $\gamma_i \in H^*(V,\bold{Z}),$ choose some generic representatives
$\Gamma_i$ of dual homology classes in $V.$ Consider an appropriate space
of triples $\{ f,C;x_1,\dots ,x_n\}$ where $C$ is a
curve of genus $g$ with $n$ marked points $x_i,$ and $f:\ C\to V$
is an algebraic map such that $f_*([C])=\beta $ and $f(x_i)\in \Gamma_i$
for all $i.$ The projection of this space to $\overline{M}_{g,n},$
under some genericity conditions, must be the cycle dual
to the cohomology class $I^V_{g,n,\beta}.$

\smallskip

As we have already remarked, the most powerful known constructions
of the classes $I^V_{g,n,\beta}$ leave the domain of algebraic
or even complex geometry, in order to satisfy the necessary
genericity assumptions. The whole subject seems to belong rather
to symplectic topology: cf. [R]. To our knowledge, only a part
of the picture of Def. 2.2 is at the moment rigorously
established even in this wider context.

\smallskip

We will use the naive picture described above in order to motivate
the expected properties of GW-classes.

\medskip

{\bf 2.3.0. Effectivity.} The meaning of this axiom is obvious.

\smallskip

{\bf 2.3.1. Grading.} The condition (2.2) expresses the following
genericity assumptions. Firstly, the space of maps $f:\ C\to V$
landing at $\beta $ must have the complex dimension of its first
order infinitesimal approximation at a point, that is
$H^0(C,f^*(N_{C/V}))$ which in turn must be unobstructed
and coincide with $\chi (C,f^*(N_{C/V})).$ By Riemann--Roch,
 $$
\chi (C,f^*(N_{C/V}))=(-K_V.\beta )+(1-g)\roman{dim}_{\bold{C}}V.
 $$

Secondly, when we constrain $f(C)$ to intersect all $\Gamma_i$'s,
this diminishes (real) dimension by $\sum_i\roman{dim}_{\bold{R}}
\Gamma_i.$

\smallskip

 From this discussion it is clear that zero-codimensional classes,
or rather numbers $\langle I^V_{g,n,\beta}\rangle$ morally count
curves constrained by incidence conditions to such a degree
that only a finite numbers of such curves occur ``generically''.
For instance, the number of curves of given degree $d$ on
$\bold{P}^2$ passing through $3d-1$ fixed points in general position and
having additional $(d-1)(d-2)/2$ double points elsewhere
must actually coincide with $\langle I^{\bold{P}^2}_{0,3d-1,d\beta_0}
\rangle (e^{\otimes 3d-1})$ where $\beta_0$ is the homology
class of a line, and $e$ the dual cohomology class of a point.
However, in more complex situations such naive counting
may be totally misleading.

\smallskip

We nevertheless use this language, but it should not be taken
too  literally.

\medskip

{\bf 2.3.2. $S_n$--covariance.} The meaning of this Axiom is
obvious. Notice that codimension zero classes are simply
$S_n$--symmetric.

\medskip

{\bf 2.3.3. Fundamental class.} The dual cycle to $e^0_V$ is
$V.$ Therefore, the l.h.s. of (2.6) imposes no constraints
on the $n$--th point of $C.$ The r.h.s. expresses this
in terms of moduli space.

\medskip

{\bf 2.3.4. Divisor.} If $I^V_{g,n,\beta}(\gamma_1\otimes \dots \otimes
\gamma_n)$ is zero--codimensional class, the same is true for
$I^V_{g,n-1,\beta}(\gamma_1 \otimes\dots\otimes \gamma_{n-1}).$
Hence the l.h.s. of (2.4) (resp. the integral at r.h.s.)
counts the number of marked curves passing through
$\Gamma_1,\dots ,\Gamma_n$ (resp. $\Gamma_1, \dots , \Gamma_{n-1}$).
But the two problems differ only by the additional presence of the
$n$--th point in the l.h.s. which may be chosen among
intersection points of $f(C)$ and $\Gamma_n.$ Their number is
$\int_{\beta}\gamma_n.$

\medskip

{\bf 2.3.5. Mapping to point.} If $\beta = f_*([C])=0,$ then $f$
maps $C$ to a point which is constrained to belong to
$\Gamma_1\cap \dots \cap \Gamma_n;$ otherwise the curve and the marked
points on it are arbitrary. This justifies the axiom (2.8) for $g=0.$

\smallskip

Unfortunately, for $g\ge 1$ this reasoning is too naive, and dealing with this
very degenerate situation requires some sophistication. In 2.4.4 below
we sketch an argument giving simultaneously (2.8)--(2.10)
and formulas for $g\ge 2.$

\medskip

{\bf 2.3.6. Splitting.} In the picture described above, the l.h.s. of
(2.11) can be represented by a cycle on $\overline{M}_{g_1,n_1+1}\times
\overline{M}_{g_2,n_2+1}$ which is the sum of cycles corresponding
to various splittings $\beta = \beta_1+\beta_2.$ When $\beta_1,\beta_2$
are chosen, we must consider pairs of maps $f_1:\ C_1\to V\times V$
and $f_2:\ C_2\to V\times V$ such that $\pi_2\circ f_1(C_1)$ and
$\pi_1\circ f_2(C_2)$ are points, $f_{i*}([C_i])=\pi^*_{3-i}(\beta_i),$
with the incidence conditions described by $\pi^*_i(\Gamma_i),$
the partition $S,$ and the additional relation
$f_1(x^{(1)}_{n_1+1})=f_2(x^{(2)}_1).$ On $V\times V,$ the latter
can be expressed by intersecting with diagonal. This leads to
the r.h.s. of (2.11).

\medskip

{\bf 2.3.7. Genus reduction.} A similar reasoning motivates (2.12).

\medskip

{\bf 2.3.8. Motivic Axiom.} For $g=0,$ and at least $V$ with
tangent sheaf $\Cal{T}_V$ generated by global sections, one can try to
construct $C^V_{0,n,\beta}$
directly as follows. Consider the scheme of maps $Map_{\beta}(\bold{P}^1,V)$
such that $f_*([\bold{P}^1])=\beta .$ Construct the morphism
 $$
a_n:\ (\bold{P}^1)^n_0\times Map_{\beta}(\bold{P}^1,V) \to
V^n\times \overline{M}_{0,n},
 $$
 $$
a_n(x_1,\dots ,x_n;f)=(f(x_1),\dots ,f(x_n);(\bold{P}^1;x_1,\dots ,x_n))
 $$
(here $(\bold{P}^1)^n_0$ is $(\bold{P}^1)^n$ with deleted diagonals.)
The closure of the image of $a_n$
is a candidate for $C^V_{0,n,\beta}.$

\smallskip

Generally, this construction is inadequate; but see 2.4 below for a
more refined version which hopefully works for certain $V.$

\smallskip

Anyway, if one takes the Motivic Axiom for granted, then
all the other axioms can and
should be stated directly in terms of $C^V_{g,n,\beta }.$
We will show how to do it for the Splitting Axiom leaving the
remaining ones to the reader.

\smallskip

For a product $\prod_{i=1}^nW_i$ and a subset $T\subset \{1,\dots ,n\},$
denote by $\pi_T:\ \prod^n_{i=1}W_i\to \prod_{i\in T}W_i$
the projection. Keeping the notation of 2.2.6, consider the following
correspondences ($\Delta =$ the diagonal class in $C^*(V\times V)$):
 $$
C_1=(id_{V^n}\times\varphi_S)^*C^V_{g,n,\beta}\in
C^*(V^n\times \overline{M}_{g_1,n_1+1}\times \overline{M}_{g_2,n_2+1}),
 $$
 $$
C_2=\sum_{\beta_1+\beta_2=\beta}\pi^*_{\{1,\dots ,n_1+2\}}
(C^V_{g_1,n_1+1,\beta_1})\pi^*_{\{n_1+3,\dots ,n+4\}}
(C^V_{g_2,n_2+1,\beta_2})\pi^*_{\{n_1+1,n_1+3\}}(\Delta )\in
$$
$$
C^*(V^{n_1+1}\times \overline{M}_{g_1,n_1+1}\times V^{n_2+1}\times
\overline{M}_{g_2,n_2+1}).
 $$
Then
 $$
C_1=\pi_{\{ 1,2\} *}t^*(C_2),
 $$
where $t\in S_{n+4}$ is the obvious reshuffling of factors.

\smallskip

We did not start with data (2.14) for two reasons. First, at present,
when constructions of GW-classes rely upon Gromov's symplectic
methods (which actually work for non-algebraic manifolds),
(2.14) looks unnecessarily restrictive. Second, using axioms
like the Splitting Axiom directly in terms of correspondences would entail a
very clumsy
notation, especially in the next section.

\medskip

{\bf 2.4. Construction project.} Fix an algebraic manifold $V$ as above.

\medskip

\proclaim{\quad 2.4.1. Definition} A stable map (to $V$) is a structure
$(C;\ \{ x_1,\dots ,x_n\} ,f)$ consisting of the following data.

\smallskip

a). $(C; x_1,\dots ,x_n)$,
or simply $C$, is a connected
reduced curve with $n\ge 0$ pairwise distinct
marked non--singular points and at most
ordinary double singular points.

\smallskip

b). $f:\ C\to V$ is a map
having no non--trivial first order infinitesimal automorphisms, identical
on $V$ and $(x_1,\dots ,x_n).$ This means that every component
of $C$ of genus 0 (resp. 1) which is contracted by $f$ must
have at least 3 (resp. 1) special (i.e., marked or singular) points
on its normalization.
\endproclaim

\medskip

For an algebraic cohomology class $\beta \in H_2(V,\bold{Z}),$
consider the stack ${\overline{M}}_{g,n}(V,\beta)$ of stable
maps to $V$ of $n$--marked curves of arithmetical genus $g$
such that $f_*([C])=\beta $ for any point $((C; x_1,\dots ,x_n),f)$
of this stack. We expect that this stack is proper and separated.

\smallskip

If $n+2g-3\ge 0,$
there is a map ${\overline{M}}_{g,n}(V,\beta )\to
V^n\times {\overline{M}}_{g,n}$ consecutively contracting the non--stable
components of $C$. For $g=0,$ it is useful to extend this map
putting ${\overline{M}}_{0,n}=${\it point} for $n\le 2.$

\medskip

\proclaim{\quad 2.4.2. Definition} $V$ is called
 convex if for
any stable map $f:\ C\to V$ of genus zero
 we have
$H^1(C,f^*(\Cal{T}_V))=0.$
\endproclaim

\smallskip

{\it Examples.} a). Generalized flag spaces $G/P$ are convex.

b). More generally, if for some $n>0,\ \Cal{T}_V^{\otimes n}$
is generated by global sections, $V$ is convex.

c). Although in general Fano manifolds are not convex
(look at exceptional curves on del Pezzo surfaces), it is
conceivable that indecomposable Fano manifolds of sufficiently large
anticanonical degree are.

\smallskip

We expect that ${\overline{M}}_{0,n}(V,\beta )$ is a smooth stack
(``an orbifold'') whenever $V$ is convex.

\medskip

{\bf 2.4.3. Construction.} For a convex $V,$ denote
by $C^V_{0,n,\beta}$ the image of the fundamental class
of ${\overline{M}}_{0,n}(V,\beta )$ in $C^*(V^n\times
{\overline{M}}_{0,n}).$

\medskip

\proclaim{\quad 2.4.4. Hope} For convex $V,$ $\{ C^V_{0,n,\beta}\}$
is a tree level motivic system of GW-classes.
\endproclaim

\smallskip

The main property to be checked is the Splitting Axiom. It must
follow from the natural structure at infinity of
${\overline{M}}_{0,n}(V,\beta )$ which is parallel to that
of ${\overline{M}}_{0,n}$ (stratification according to the
degeneration graph of a curve).

\smallskip

If $g\ge 1$ and/or $V$ is not convex,  the fundamental
class of ${\overline{M}}_{g,n}(V,\beta )$ is ``too big'', and
$C^V_{g,n,\beta }$ must be defined as image of a characteristic
class of an appropriate ``obstruction complex'' on
this moduli space. We illustrate the arising complications
on the ``mapping to point'' example.

\medskip

{\bf 2.4.5. Mapping to point.} By Definition 2.4.1, we have
 $$
{\overline{M}}_{g,n}(V,0)={\overline{M}}_{g,n}\times V.
 $$
The complex dimension of this space is bigger than the expected one
by $g\roman{dim}_{\bold{C}}V:=G.$ This is precisely the
rank of the locally free obstruction sheaf $\Cal{T}^{(1)}$
on ${\overline{M}}_{0,n}(V,\beta )$
whose geometric fiber at the point $[f],\ f(C)=v \in V,$
is
 $$
H^1(C,f^*(\Cal{T}_V))\cong H^1(C,\Cal{O}_C)\otimes T_vV.
 $$
Denote by $\Cal{E}_{g,n}$ the locally free sheaf $R^1\pi_*\Cal{O}$
on ${\overline{M}}_{g,n}$ where $\pi$ is the projection of the
universal curve. We have $\Cal{T}^{(1)}\cong \Cal{E}_{g,n}\boxtimes
\Cal{T}_V.$

\smallskip

Notice that $\Cal{E}_{g,n}$ is simply the pullback of one of the basic
sheaves $\Cal{E}_{0,3}\cong 0,\ \Cal{E}_{1,1},\ \Cal{E}_{g,0},$
for $g=0,1,\ge 2$ respectively.

\smallskip

Consider now the Euler class $c_G(\Cal{T}^{(1)})$ and denote by
$p_1,p_2$ the two projections of
${\overline{M}}_{g,n}\times V$.
An intuitive argument appealing to our desire to
pass to an unobstructed situation suggests the following definition:
 $$
I^V_{g,n,0}(\gamma_1\otimes\cdots\otimes\gamma_n)=
p_{1*}(c_G(\Cal{E}\boxtimes\Cal{T}_V)\wedge p_2^*(
\gamma_1\wedge\dots\wedge\gamma_n)).
 $$

\smallskip

We hope to develop this construction in a separate
publication.

\smallskip

We will end our general discussion of GW-classes with the following
two constructions.

\medskip

{\bf 2.5. Tensor product.} Let $\{ I^{V_i}_{g,n,\beta}\},\ i=1,2,$
be two full (or only tree level) GW-systems for $V_1,V_2,\ H^1(V_i)=0.$
Put
 $$
I^{V_1\times V_2}_{g,n,\beta_1\oplus\beta_2}
((\gamma^{(1)}_1\otimes\gamma^{(2)}_1)\otimes \dots \otimes
(\gamma^{(1)}_n\otimes\gamma^{(2)}_n)):=
 $$
 $$
\varepsilon I^{V_1}_{g,n,\beta_1}(\gamma^{(1)}_1\otimes \dots
\otimes\gamma^{(1)}_n)
\wedge      I^{V_2}_{g,n,\beta_2}(\gamma^{(2)}_1\otimes \dots
\otimes\gamma^{(2)}_n),
\eqno(2.16)
 $$
where $\varepsilon$ is the sign of the relevant permutation
of odd-dimensional arguments.

\medskip

{\bf 2.5.1. Claim.} (2.15) is a full (resp. tree level) system
of GW-clases for $V_1\times V_2$ which is called {\it the tensor
product} of given systems.

\medskip

In fact, one easily checks all axioms, including the refined version
of the Mapping to Point Axiom.

\smallskip

Notice that even if one is interested only in codimension zero
classes of the tensor product, one has to know all classes of
the factors. In the tree level setting, they can
be in turn be calculated from the codimension zero classes of the factors,
but in a highly non-trivial way. In fact, we have:

\medskip

\proclaim{\quad 2.5.2. Proposition} Let a tree level system
of GW-classes $I^V_{0,n,\beta}$ be given for $V.$ Then it
can be uniquely reconstructed from its codimension zero
subsystem.
\endproclaim

\smallskip

{\bf Proof.} In fact, consider a class of codimension $\ge 1$
with $n\ge 4.$ The Splitting Axiom (2.11) allows one to calculate
its restrictions to all boundary components of the moduli
space (corresponding to lesser values of $n$). It remains
to show that $\cap_S \roman{Ker}\ \varphi^*_S=H^{2n-6}({\overline{M}}_{0,n}).$
In fact, let $d_S\in H^2({\overline{M}}_{0,n})$ be the dual
class of the boundary component corresponding to the partition $S.$
The whole cohomology ring is generated by these classes
(see [Ke]). On the other hand, $\varphi_{S*}\varphi_{S}^*(\alpha )=
\alpha \wedge d_S$ for any class $\alpha .$ And if
$\alpha \ne 0,\ |\alpha |< 2(n-3),$ then by Poincar\'e duality
there exists some non-constant monomial $d$ in $d_S$ such that
$\int_{{\overline{M}}_{0,n}}\alpha\wedge d\ne 0.$ Hence
$\alpha \notin \cap_S\roman{Ker}\ \varphi^*_S.$

\smallskip

Summarizing,  we may say that
the tensor product can be defined on the codimension zero
subsystems, but there are no simple formulas for doing it.

\smallskip

{\bf 2.5.3. Cusp classes.} Generally, let us call {\it cusp classes}
those elements of $H^*(\overline{M}_{g,n})$ which vanish on all boundary
divisors of this moduli space. For $g\ge 1,$ non-trivial
cusp classes may exist (e.g. the Ramanujan class for $g=1,\ n=11$).
It would be interesting to have examples of GW-classes
with non-trivial cusp components.

\medskip

{\bf 2.6. Restricted GW-systems.} Let $C^*$ be an intersection theory
such that a given GW-system $\{I^V_{g,n,\beta}\}$ can be represented
by $C$--correspondences. We have $C^*(V^n)\subset H^*(V)^{\otimes n}.$
We will say that the maps $I^V_{g,n,\beta}$ restricted to $C^*(V^n)$
form {\it the restricted GW-system.} Slightly elaborating the discussion
of the Splitting Axiom in 2.2.8, one can convince oneself that
all the axioms restricted to $\oplus_nC^*(V^n)$ make sense
and can be stated entirely in terms of this restriction, without appealing
to extra cohomology classes like $\Delta_a$ in (2.11) and (2.12).

\smallskip

This is useful for those enumerative problems where we want to consider
incidence conditions stated in terms of algebraic cycles only.

\newpage


\centerline{\bf \S 3. First Reconstruction Theorem}

\bigskip

\proclaim{\quad 3.1. Theorem} Let $V$ be a manifold for which a tree
level system of GW-classes $\{ I^V_{0,n,\beta}\}$ exists.

If $H^*(V)$ as a ring is generated by $H^2(V),$ then $\{ I^V_{0,n,\beta}\}$
can be uniquely reconstructed starting with the following system
of codimension zero basic classes:
 $$
\{ I^V_{0,3,\beta}(\gamma_1\otimes\gamma_2 \otimes \gamma_3)\ |\
(-K_V.\beta )\le 2\roman{dim}_{\bold{C}}V+1;
 $$
 $$
\sum_{i=1}^3|\gamma_i|=2(-K_V.\beta )+2\roman{dim}_{\bold{C}}V;\
|\gamma_3|=2\}. \eqno(3.1)
 $$
\endproclaim

\medskip

{\bf 3.1.1. Comments.} We may and usually will choose $|\gamma_i|$
from the elements of a fixed basis of $H^*(V).$ Then, if $V$ is Fano,
(3.1) is a finite set because the degree of $\beta$ is also bounded.
For instance, if $V=\bold{P}^n,$ (3.1) is satisfied only by $\beta =0$
and $\beta$ = class of a line. The $\beta = 0$ case is settled
by (2.8). For the line, (3.1) gives $\{ \gamma_1,\gamma_2,\gamma_3\} =
\{ e^{2n},e^{2n},e^{2}\} ,$ where $e^{2i}=c_1(\Cal{O}(1))^i.$
Since $e^{2n}$ is the dual class of a point, one can imaginatively
say that all enumerative problems about rational curves in $\bold{P}^n$
eventually reduce to counting the number of lines passing through two points.

\smallskip

On the other hand, for Calabi--Yau varieties with $K_V=0$ (3.1)
does not restrict $\beta$ at all. Besides, $H^2(V)$ almost never
generates $H^*(V).$ Nevertheless, Theorem 3.1 does say something
about this case as well.

\smallskip

a). The algebraic (or Hodge) part of cohomology may be generated
by $H^2$, and the corresponding restricted GW-system can be
reconstructed from (3.1).

\smallskip

b). Then Theorem 3.1 says that all tree level classes with algebraic
arguments can be reconstructed if one knows all $\beta$--contributions
to the triple quantum intersection indices. This information
is conjecturally supplied by the Mirror family.

\medskip

{\bf 3.2. Proof.} It will consist of several reduction steps.

\smallskip

{\bf 3.2.1. Step 1.} {\it Every class} $I^V_{0,n,\beta}(\gamma_1
\otimes \dots \otimes\gamma_n)$ {\it of codimension} $\ge 1$
{\it with} $n\ge 4$ {\it can be reconstructed from classes with lesser
values of} $n.$

\smallskip

The proof of the Proposition 2.5.2 shows this.

\smallskip

It remains to deal with codimension zero classes, that is,
numbers
 $$
\langle I^V_{0,n,\beta}\rangle (\gamma_1\otimes \dots \otimes \gamma_n),\
\sum_{i=1}^n|\gamma_i|=2(-K_V.\beta)+2\roman{dim}_{\bold{C}}V+2(n-3).
\eqno(3.2)
 $$
We start with some preliminaries.

\medskip

{\bf 3.2.2. Quadratic relations.} Fix some $\{ \beta ;\gamma_1,\dots
\gamma_N\} ,\ N\ge 4,$ and four pairwise distinct indices
$\{ i,j,k,l\} \subset \{1,\dots ,N\} .$ Assume that
$I^V_{0,N,\beta}(\gamma_1\otimes \dots \otimes \gamma_N)$
has topological codimension two. Take the cup product of this class
with the following linear relation between the generators $d_S$
established in [Ke]:
 $$
\sum_{\{ ijSkl\}}d_S=\sum_{\{ ikTjl\}}d_T,
 $$
where $\sum_{\{ ijSkl\}}$ means that we sum over all partitions $S$ such that
$i,j\in S_1,\ k,l\in S_2,$ or vice versa.

Calculate the degrees of all summands using (2.11). We get the following
fundamental system of quadratic relations among codimension zero classes:
 $$
\sum_{\{ ijSkl\}}\sum_{\beta_1+\beta_2=\beta}\sum_{a,b}\varepsilon (S)
\langle I^V_{0,|S_1|+1,\beta_1}\rangle ((\otimes_{r\in
S_1})\gamma_r)\otimes\Delta_a)
g^{ab}\langle I^V_{0,|S_2|+1,\beta_2}\rangle (\Delta_b\otimes
(\otimes_{s\in S_2}\gamma_s))=
$$
$$
\sum_{\{ ikTjl\}}\sum_{\beta_1+\beta_2=\beta}\sum_{a,b}\varepsilon (T)
\langle I^V_{0,|T_1|+1,\beta_1}\rangle ((\otimes_{r\in
T_1})\gamma_r)\otimes\Delta_a)
g^{ab}\langle I^V_{0,|T_2|+1,\beta_2}\rangle (\Delta_b\otimes
(\otimes_{s\in T_2}\gamma_s)).
\eqno(3.3)
$$

Now, define a partial order on pairs $(\beta ,n),\ \beta\in B,\ n\ge 3,$
by setting $(\beta ,n)>(\beta^{\prime},n^{\prime}),$ iff either
$\beta =\beta^{\prime}+\beta^{\prime\prime},\ \beta^{\prime},
\beta^{\prime\prime}\in B,\ \beta^{\prime\prime}\ne 0,$ or
$\beta =\beta^{\prime},\ n>n^{\prime}.$

\smallskip

Observe that the highest order terms enter in (3.3) linearly.
In fact, for these terms we have either $\beta_1=\beta$ or
$\beta_2=\beta .$ The complementary class, with
$\beta_2=0$ (resp. $\beta_1=0$), can be non-zero only if
$|S_2|=2$ or $|T_2|=2$ (resp. $|S_1|=2$ or $|T_1|=2$): see (2.8).
Hence there are four possibilities: $S_1=\{ i,j\} ; S_2=\{ k,l\} ;
T_1=\{ i,k\} ; T_2=\{ j,l\} .$

\smallskip

Let us look, say, at the first group of highest terms:
  $$
\varepsilon (S)\sum_{a,b}
\langle I^V_{0,3,0}\rangle (\gamma_i\otimes\gamma_j\otimes \Delta_a)
g^{ab}\langle I^V_{0,n-1,\beta}\rangle (\Delta_b\otimes
(\otimes_{s\ne i,j}\gamma_s)).
\eqno(3.4)
  $$
We have by (2.8):
  $$
\langle I^V_{0,3,0}\rangle (\gamma_i\otimes\gamma_j\otimes\Delta_a)=
\int_V\gamma_i\wedge\gamma_j\wedge\Delta_a.
  $$
Since $\langle I^V_{0,N-1,\beta}\rangle$ is (poly)linear,
we can rewrite (3.4) as
  $$
\varepsilon (S)\langle I^V_{0,N-1,\beta}\rangle
(\sum_{a,b}(\int_V\gamma_i\wedge\gamma_j\wedge\Delta_a)
g^{ab}\Delta_b\otimes (\otimes_{s\ne i,j}\gamma_s))=
  $$
  $$
\varepsilon (S)\langle I^V_{0,N-1,\beta}\rangle
(\gamma_i\wedge\gamma_j\otimes (\otimes_{s\ne i,j}\gamma_s)).
\eqno(3.5)
  $$
Using analogs of (3.5) for all four groups of highest order
terms we can finally write (3.3) as

\pagebreak

  $$
\pm \langle I^V_{0,N-1,\beta}\rangle (\gamma_i\wedge\gamma_j\otimes
(\otimes_{s\ne i,j}\gamma_s))
\pm \langle I^V_{0,N-1,\beta}\rangle (\gamma_k\wedge\gamma_l
\otimes (\otimes_{s\ne k,l}\gamma_s ))
  $$
  $$
\pm \langle I^V_{0,N-1,\beta}\rangle (\gamma_i\wedge\gamma_k\otimes
(\otimes_{s\ne i,k}\gamma_s))
\pm \langle I^V_{0,N-1,\beta}\rangle (\gamma_j\wedge\gamma_l\otimes
(\otimes_{s\ne j,l}\gamma_s))=
  $$
  $$
\roman{a\ quadratic\ combination\ of\ lower\ order\ terms.}
\eqno(3.6)
  $$

\medskip

 {\bf 3.2.3. Step 2.} {\it Every class} $I^V_{0,n,\beta}(\gamma_1
\otimes \dots \otimes\gamma_n)$ {\it of codimension zero with}
$n\ge 4$ {\it can be reconstructed from basic classes (with}
$n=3).$

\smallskip

In fact, it suffices to calculate numbers $\langle I^V_{0,n,\beta}\rangle
(\gamma_1\otimes \cdots \otimes\gamma_n)$ for $n\ge 4$ and
  $$
2\roman{dim}_{\bold{C}}(V)\ge |\gamma_1|\ge\dots \ge |\gamma_n|\ge 4
  $$
(if $|\gamma_n|=2$, we can apply the Divisor Axiom to reduce $n$).
We will now for the first time use the assumption that $H^*(V)$
is generated by $H^2(V)$ and write $\gamma_n=\sum_i\delta_i\wedge
\delta_i^{\prime}$ for some $\delta_i,\delta_i^{\prime}$ with
$|\delta_i^{\prime}|=2.$ Clearly, it suffices to treat the case
$\gamma_n=\delta\wedge\delta^{\prime}, |\delta^{\prime}|=2.$
Apply the construction of 3.2.2 to the codimension two class
  $$
I^V_{0,n+1,\beta}(\gamma_1\otimes \dots \otimes \gamma_{n-1}
\otimes\delta\otimes\delta^{\prime})
  $$
and indices $\{ i,j,k,l\} =\{ 1,2,n,n+1\}.$ Relation  (3.6) becomes
  $$
\pm\langle I^V_{0,n,\beta}(\gamma_1\wedge\gamma_2\otimes\gamma_3\otimes
\dots\otimes\gamma_{n-1}\otimes\delta\otimes\delta^{\prime})\pm
\langle I^V_{0,n,\beta}\rangle(\gamma_1\otimes\dots\otimes\gamma_{n-1}
\otimes\delta\wedge\delta^{\prime})
  $$
  $$
\pm\langle I^V_{0,n,\beta}\rangle (\gamma_1\wedge\delta\otimes\gamma_2
\otimes\dots\otimes\gamma_{n-1}\otimes\delta^{\prime})
\pm \langle I^V_{0,n,\beta}\rangle (\gamma_1\otimes\gamma_2\wedge
\delta^{\prime}\otimes\gamma_3\otimes\dots\otimes\gamma_{n-1}\otimes
\delta )=
  $$
  $$
\roman{a\ quadratic \ combination\ of\ lower\ terms}.
\eqno(3.7)
  $$

Now, the second summand in (3.7) is our initial class
$\langle I^V_{0,n,\beta}\rangle (\gamma_1\otimes\dots\otimes
\gamma_{n-1}\otimes\gamma_n).$ The first and the third
summands are lifted from $\overline{M}_{0,n-1}$because of the
Divisor Axiom. Finally, in the fourth summand
the last argument is of lesser dimension than in the initial class:
$|\delta |<|\gamma_n|.$ If $|\delta |=2,$ it is lifted from
$\overline{M}_{0,n-1};$ if $|\delta |>2,$ we can repeat the
same trick applying it to this summand. In a finite number of
iterations, we will reduce $n.$

\medskip

{\bf 3.2.4. Step 3.} {\it Every basic class} $I^V_{0,3,\beta}$
{\it can be calculated via those with restrictions (3.1).}

\smallskip

In fact, if $|\gamma_3|\ge 4$ in $I^V_{0,3,\beta}(\gamma_1\otimes
\gamma_2\otimes\gamma_3),\ \beta \ne 0,$ (the $\beta =0$ case
is given by (2.8)), then we can apply to this class the
reduction procedure described above and diminish $|\gamma_3|.$
The remaining conditions follow from the Grading Axiom.

\newpage


\centerline{\bf \S 4. Potential, associativity relations,}
\centerline{\bf and quantum cohomology}

\bigskip

{\bf 4.1. Setup.} let $M$ be a supermanifold endowed with a tensor $g$ of rank
two and a tensor $A$ of rank three. To fix our sign conventions, it
is convenient to choose a (local) coordinate system $\{ x^a\}$
which defines the basis $\partial_a=\partial /\partial x^a$
of vector fields and the basis $dx^a$ of 1-forms. Our tensors
then have components $g_{ab}$ and $A_{ab}^c.$

\smallskip

Generally, $\tilde{x}$ denotes the $\bold{Z}/2\bold{Z}$--degree of $x.$
To simplify notation, in superscripts we replace $\tilde{x}_a$ by $a$
so that e.g. $(-1)^{\tilde{x}_a\tilde{x}_b+\tilde{x}_c\tilde{x}_d}$
becomes $(-1)^{ab+cd}.$ Hopefully, this will not lead to a confusion.

\smallskip

We want $g$ and $A$ to be even, i.e. $\tilde{g}_{ab}=(-1)^{a+b},\
\tilde{A}_{ab}^c=(-1)^{a+b+c}.$ The sign conventions about the
de Rham complex are: $d$ is odd, and $\Omega^*_M$ is the symmetric
algebra of $\Omega^1_M.$

\medskip

{\bf 4.2. Pairing, multiplication and connection (Dubrovin's
formalism).} We use $g_{ab}$ in order to define an even
$\Cal{O}_M$--pairing on the tangent sheaf $\Cal{T}_M:$
  $$
\langle \partial_a,\partial_b\rangle :=g_{ab}.
\eqno(4.1)
  $$
We will always assume it to be symmetric ($g_{ba}=(-1)^{ab}g_{ab}$)
and non-degenerate, so that the inverse matrix $(g^{ab})=(g_{ab})^{-1}$
exists. In addition, we will assume that $g_{ab}$ are constant,
so that it defines a flat metric in a flat coordinate system
(for non--flat coordinates, cf. [D]).

\smallskip

We use $A_{ab}^c$ in order to construct, firstly, a structure
of $\Cal{O}_M$--algebra on $\Cal{T}_M$ with multiplication
$\circ :$
  $$
\partial_a\circ\partial_b:=\sum_cA_{ab}^c\partial_c
\eqno(4.2)
  $$
and secondly, a family of connections on $\Cal{T}_M$ depending on a even
parameter $\lambda$ and defined by the covariant differential
  $$
\nabla_{\lambda}(\partial_b):=
\lambda \sum_{a,c}dx^aA_{ab}^c\otimes\partial_c
\eqno (4.3)
  $$
or equivalently, by covariant derivatives
  $$
\nabla_{\lambda ,\partial_a}(\partial_b):=
\lambda\sum_cA_{ab}^c\partial_c=\lambda (\partial_a\circ\partial_b).
\eqno(4.4)
  $$

We will now consecutively impose some relations upon $A,g,$ and
interpret them both in terms of multiplications and connections.

\medskip

{\bf 4.2.1. Commutativity/vanishing torsion.}
  $$
\forall a,b,c,\qquad A_{ba}^c=(-1)^{ab}A_{ab}^c.
\eqno(4.5)
  $$
In view of (4.2), this means supercommutativity of $(\Cal{T}_M,\circ ).$
 From (4.4) it follows that
  $$
\forall a,b,\qquad \nabla_{\lambda ,\partial_a}(\partial_b)=
(-1)^{ab}\nabla_{\lambda ,\partial_b}(\partial_a).
\eqno(4.6)
  $$

\medskip

{\bf 4.2.2. Associativity/flatness.}
  $$
\forall a,b,c,d, \qquad \sum_eA_{ab}^eA_{ec}^d=
(-1)^{a(b+c)}\sum_eA_{bc}^eA_{ea}^d,
\eqno(4.7a)
  $$
  $$
\forall a,b,c,d,\qquad \partial_dA_{ab}^c=(-1)^{ad}\partial_aA_{db}^c.
\eqno(4.7b)
  $$

Using (4.2) one checks that (4.7a) is equivalent to the associativity
relations $(\partial_a\circ\partial_b)\circ\partial_c=
\partial_a\circ (\partial_b\circ\partial_c).$

\smallskip

Using (4.3) one checks that (4.7a) and (4.7b) {\it together}
are equivalent to $\nabla_{\lambda}^2=0.$ More precisely,
if one puts symbolically $\nabla_{\lambda}=\nabla_0+\lambda A,$
then (4.7b) is $\nabla_0(A)=0,$ and (4.7a) is equivalent
to $[A,A]=0$ if one assumes (4.5) or (4.6).

\medskip

{\bf 4.2.3. Frobenius algebra.}
Put $A_{abc}:=\sum_a A_{ab}^eg_{ec}.$ The next relation we impose is:
  $$
\forall a,b,c,\qquad A_{abc}=(-1)^{a(b+c)}A_{bca}.
\eqno(4.8)
  $$
Together with (4.6), this means that $A_{abc}$ is $S_3$--invariant
(in the sense of superalgebra).

\smallskip

In terms of multiplication, (4.8) reads
  $$
\langle \partial_a\circ\partial_b,\partial_c\rangle =
\langle \partial_a,\partial_b\circ\partial_c\rangle ,
  $$
that is, the scalar product is invariant wrt multiplication.
In terms of connection, (4.8) reads (use (4.4) and (4.1)):
  $$
\langle \nabla_{\lambda ,\partial_a}(\partial_b),\partial_c\rangle =
(-1)^{ab}\langle \partial_b,\nabla_{\lambda ,\partial_a}(\partial_c)\rangle
\eqno(4.9)
  $$

\medskip

{\bf 4.2.4. Identity.} Assume that the coordinate vector field
$\partial_0$ is even, and
  $$
\forall b,c,\qquad A_{0b}^c=\delta_{bc}
\eqno(4.10a)
  $$
or equivalently,
  $$
A_{0bc}=g_{bc}. \eqno(4.10b)
  $$
According to (4.2), this means that $\partial_0$ is identity
in $(\Cal{T}_M,\circ ).$ According to (4.3), this is equivalent to
  $$
\nabla_{\lambda}(\partial_0)=\lambda\sum_adx^a\partial_a
  $$
or more suggestively,
  $$
\nabla_{\lambda}(\partial_0)=\lambda d.
  $$
This also means that
  $$
\forall b,\qquad \nabla_{\lambda ,\partial_0}(\partial_b)=\lambda
\partial_b .
\eqno(4.11)
  $$

\medskip

{\bf 4.2.5. Potential.} An even (local) function $\Phi$ on $M$
is called a {\it potential} for the $(A,g)$--structure, if
  $$
\forall a,b,c,\qquad A_{abc}=\partial_a\partial_b\partial_c\Phi.
\eqno(4.12)
  $$

Such a potential always exists locally.
On the other hand, for any function
$\Phi $ and the tensor of its derivatives $A_{abc},$
the $S_3$--invariance, in particular (4.8), is automatic.
If we then define $A_{ab}^c$ by $\sum_eA_{abe}g^{ec},$ then
(4.5) is also automatic. If in addition $g^{ec}$ are constant,
(4.7b) follows.

\smallskip

The crucial associativity relations (4.7a) then become a
remarkable system of quadratic differential equations called
WDVV--equations in [D]:
  $$
\forall a,b,c,d,\qquad \sum_{ef}\partial_a\partial_b\partial_e\Phi\cdot
g^{ef}\partial_f\partial_c\partial_d\Phi =
(-1)^{a(b+c)}\sum_{ef}\partial_b\partial_c\partial_e\Phi \cdot
g^{ef}\partial_f\partial_a\partial_d\Phi .
\eqno(4.13)
  $$

We will now show how to derive a potential $(A,g)$--structure
from a tree level system of GW-classes.

\medskip

{\bf 4.3. GW-potential.} Let $V$ be a manifold equipped with a
system of tree level GW-classes. We will actually use only
the numbers $\langle I^V_{0,n,\beta}\rangle (\gamma_1\otimes\dots
\otimes\gamma_n)$ with properties postulated in \S 2.

\smallskip

We will consider $H^*(V,\bold{C})$ as a linear superspace
$\bold{Z}/2\bold{Z}$--graded by $\tilde{\gamma}:=|\gamma |
\roman{mod}\ 2,$ and as a supermanifold which we then denote
$H^V.$ Our potential $\Phi_{\omega}$ will depend on a choice
of a class $\omega\in H^2(V,\bold{C})$ whose real part lies
in the K\"ahler cone. We first define $\Phi_{\omega}$
as a formal sum depending on a variable point $\gamma\in H^V:$
  $$
\Phi_{\omega}(\gamma):=\sum_{n\ge 3}\sum_{\beta}
e^{-\int_{\beta}\omega }\frac{1}{n!}\langle I^V_{0,n,\beta}\rangle
(\gamma^{\otimes n}).
\eqno(4.14)
  $$

To make sense of this expression, choose a basis $\{ \Delta_a\}$ of
$H^*(V,\bold{C}),$ write the generic point as $\gamma =\sum_{a=0}^D
x^a\Delta_a,\ \tilde{x}^a=\tilde{\Delta}_a,$ and define the metric
by Poincar\'e duality:
  $$
\langle \partial_a,\partial_b\rangle =g_{ab}:=\int_V
\Delta_a\wedge\Delta_b.
\eqno(4.15)
  $$
Since $\langle I^V_{0,n,\beta}\rangle (\gamma_1\otimes\dots\otimes\gamma_n)$
is simply $S_n$--invariant when all $\gamma_i$ are even,
we have
  $$
\frac{1}{n!}\langle I^V_{0,n,\beta}\rangle (\gamma^{\otimes n})=
  $$
  $$
\sum_{n_0+\dots +n_D=n\ge 3}\frac{\varepsilon(n_0,\dots ,n_D)}{n_1!\dots n_D!}
\langle I^V_{0,n,\beta}\rangle
(\Delta_0^{\otimes n_0}\otimes\dots\otimes \Delta_D^{\otimes n_D})
x_0^{n_0}\dots x_D^{n_D},
\eqno(4.16)
  $$
where $\varepsilon (n_0,\dots ,n_D)=\varepsilon =\pm 1$ is the sign
acquired in the supercommutative algebra \linebreak $S(H^*(V))[x^0,\dots ,x^D]$
after reshuffling $\prod_{a=0}^D(x^a\Delta_a)^{n_a}=\varepsilon
\prod_{a=0}^D\Delta_a^{n_a}\prod_{a=0}^D(x^a)^{n_a}.$

\smallskip

There are several natural convergence assumptions that can be made
about (4.14).

\medskip

A. $\forall n\ge 3,$ {\it there exists only finitely many effective}
$\beta$ {\it satisfying the grading condition (3.2) for the
zero--codimensional classes.}

\smallskip

This is the case of Fano manifolds. If this hypothesis is satisfied,
(4.14) can be interpreted as a formal series in $x^a$

\newpage

  $$
\Phi_{\omega}(\gamma )=
  $$
  $$
\sum_{n_0+\dots +n_D\ge 3}\sum_{\beta}
\frac{\varepsilon (n_0,\dots ,n_D)e^{-\int_{\beta}\omega} }
{n_0!\dots n_D! }
\langle I^V_{0,n,\beta}\rangle
(\Delta_0^{\otimes n_0}\otimes \dots\otimes\Delta_D^{\otimes n_D})
(x^0)^{n_0}\dots (x^D)^{n_D},
\eqno(4.17)
  $$
because each interior sum $\sum_{\beta}$ is effectively finite.

\medskip

B. {\it The previous condition is not satisfied, but each}
$\sum_{\beta}$ {\it in the r.h.s. of (4.17) converges for each} $n,$
{\it at least when} $\omega$ {\it has a sufficiently large real
K\"ahler part.}

\smallskip

(Conjecturally, this is the case for Calabi--Yau manifolds).

\smallskip

Then again, (4.17) is a well-defined formal series.

\medskip

C. A {\it or} B {\it is satisfied, and in addition} $\Phi_{\omega}(\gamma )$
{\it converges in a subdomain} $M$ {\it of} $H^V$ {\it (which may depend
on} $\omega$) {\it as a function of} $\{ x^a\} .$

\smallskip

We expect this to be generally true, because of the conjectural
exponential growth estimates for $\frac{1}{n!}\langle I^V_{0,n,\beta}\rangle .$

\smallskip

The following formal calculations can be easily justified in each of
these contexts.

\medskip

\proclaim{\quad 4.4. Proposition} Let $\gamma =\gamma^{\prime}+
\gamma^2 +\gamma^0,$ where $|\gamma^2|=2,\ |\gamma^0|=0,$
and $\gamma^{\prime}=$ the sum of components of dimension $\ne 0,2.$
Then

a). $\Phi_{\omega}(\gamma^{\prime}+\gamma^2+\gamma^0)=
\Phi_{\omega}(\gamma^{\prime}+\gamma^2)+$
a function quadratic in $(\gamma^{\prime}+\gamma^2).$

b). If $x^0$ is the coefficient of $\gamma$ at $\Delta_0=e^0_V$
(identity in $H^*(V)$) so that $\gamma^0=x^0\Delta_0,$ we have
  $$
\partial_0\partial_b\partial_c\Phi_{\omega}(\gamma )= g_{bc}.
\eqno(4.18)
  $$

c). We have
  $$ \Phi_{\omega}(\gamma^{\prime}+\gamma^2)=
\Phi_{\omega -\gamma^2 }(\gamma^{\prime})+
  $$
  $$
{\ a\ function\ quadratic\ in\ }
\gamma ,\ \gamma^{\prime}. \eqno(4.19)
  $$

\endproclaim

\smallskip

{\bf Proof.} a). $\gamma^{\otimes n} = (\gamma^{\prime}+\gamma^2)^{\otimes n}
+\sum_{i+j=n-1}(\gamma^{\prime}+\gamma^2)^i\otimes\gamma^0\otimes
(\gamma^{\prime}+\gamma^2)^j$ plus terms containing $\gamma^0$ at least twice.
 From the Fundamental Class Axiom it follows that $\langle I^V_{0,n,\beta}
\rangle =0$
for $n\ge 4,$ if $e^0_V$ is among the arguments. Hence the contribution
of $\gamma^0$ to $\Phi_{\omega}(\gamma )$ is restricted to the terms
$n=3$ in (4.17), and they are (no more than) quadratic in
$\gamma^{\prime}+\gamma^2.$

\smallskip

b). We can now calculate $\partial_0\partial_b\partial_c\Phi_{\omega}$
taking into account only the $n=3$ terms:
  $$
\partial_0\partial_b\partial_c\langle I^V_{0,3,\beta}\rangle
(\gamma\otimes\gamma\otimes\gamma )=6\langle I^V_{0,3,\beta}\rangle
(e^0\otimes\Delta_b\otimes\Delta_c).
  $$
This vanishes if $\beta \ne 0$ and is $g_{bc}$ otherwise: see (2.7).

\smallskip

c). We have
  $$
\Phi_{\omega -\gamma^2}(\gamma^{\prime})=\sum_{p\ge 3}\sum_{\beta}
e^{-\int_{\beta}(\omega - \gamma^2) }\frac{1}{p!}
\langle I^V_{0,p,\beta}\rangle (\gamma^{\prime\otimes p}) =
  $$
  $$
\sum_{p\ge 3,q\ge 0}\sum_{\beta}e^{-\int_{\beta}\omega }
\frac{1}{p!q!}\langle I^V_{0,p,\beta}\rangle
(\gamma^{\prime\otimes p})(\int_{\beta}\gamma^2)^q;
  $$
  $$
\Phi_{\omega}(\gamma^{\prime}+\gamma^2)=\sum_{n\ge 3}\sum_{\beta}
e^{-\int_{\beta}\omega }\frac{1}{n!}\langle I^V_{0,n,\beta}\rangle
((\gamma^{\prime}+\gamma^2)^{\otimes n})=
  $$
  $$
\sum_{p+q\ge 3}\sum_{\beta}e^{-\int_{\beta}\omega }\frac{1}{p!q!}
\langle I^V_{0,n,\beta}\rangle (\gamma^{\prime\otimes p}\otimes
(\gamma^2)^{\otimes q}).
  $$
Finally, because of the Divisor Axiom, for $p \ge 3,\ q\ge 0,$
we have
  $$
\langle I^V_{0,p+q,\beta}\rangle (\gamma^{\prime \otimes p}\otimes
(\gamma^2)^{\otimes q})=
\langle I^V_{0,p,\beta}\rangle (\gamma^{\prime \otimes p})
(\int_{\beta}\gamma^2)^q.
  $$

\medskip

\proclaim{\quad 4.5. Theorem-Definition} The tensors $g_{ab}$ and
$\partial_a\partial_b\partial_c\Phi_{\omega}$ define a potential
Dubrovin structure satisfying all the properties (4.1)--(4.11),
with $\partial_0$ as identity.

The fibers of $\Cal{T}_M$ endowed with multiplication $\circ$
are called the quantum cohomology rings of $V$ associated
with the tree level system of GW-classes $I^V.$

\endproclaim

\smallskip

{\bf Proof.} It remains only to check the relations (4.13).
Let us calculate the l.h.s. of (4.13) using (4.17). The terms
with fixed $e,f$ are:
  $$
\sum_{n_1\ge 3,\beta_1}e^{-\int_{\beta_1}\omega }\frac{1}{(n_1-3)!}
\langle I^V_{0,n_1,\beta_1}\rangle
(\gamma^{\otimes (n_1-3)}\otimes\Delta_a\otimes\Delta_b\otimes
\Delta_e)g^{ef}\times
  $$
  $$
\times
\sum_{n_2\ge 3,\beta_2}e^{-\int_{\beta_2}\omega}\frac{1}{(n_2-3)!}
\langle I^V_{0,n_2,\beta_2}\rangle
(\Delta_f\otimes\Delta_c\otimes\Delta_d\otimes\gamma^{\otimes (n_2-3)})=
  $$
  $$
\sum_{n\ge 6,\beta}\frac{1}{(n-6)!}e^{-\int_{\beta}\omega }
\sum_{\beta_1+\beta_2=\beta}\sum_{n_1+n_2=n}
{{n-6}\choose{n_1-3}}\langle I^V_{0,n_1,\beta_1}\rangle
(\gamma^{\otimes (n_1-3)}\otimes\Delta_a\otimes\Delta_b\otimes
\Delta_e)g^{ef}\times
$$
$$
\times\langle I^V_{0,n_2,\beta_2}\rangle
(\Delta_f\otimes\Delta_c\otimes\Delta_d\otimes\gamma^{\otimes (n_2-3)}).
  $$
Rewriting similarly the r.h.s., we see that (4.13) is equivalent to
the family of identities
  $$ \sum_{n_1+n_2=n}\sum_{\beta_1+\beta_2}\sum_{e,f}
{{n-6}\choose{n_1-3}}\langle I^V_{0,n_1,\beta_1}\rangle
(\gamma^{\otimes
(n_1-3)}\otimes\Delta_a\otimes\Delta_b\otimes\Delta_e)g^{ef}\times
  $$
  $$
\times\langle I^V_{0,n_2,\beta_2}\rangle
(\Delta_f\otimes\Delta_c\otimes\Delta_d\otimes
\gamma^{\otimes (n_2-3)})=
  $$
  $$
(-1)^{a(b+c)}\sum_{n_1+n_2=n}\sum_{\beta_1+\beta_2=\beta}\sum_{e,f}
{{n-6}\choose{n_1-3}}\langle I^V_{0,n_1,\beta_1}\rangle
(\gamma^{\otimes (n_1-3)}\otimes\Delta_b\otimes\Delta_c\otimes\Delta_e)
g^{ef}\times
  $$
  $$
\times
\langle I^V_{0,n_2,\beta_2}\rangle
(\Delta_f\otimes\Delta_a\otimes\Delta_d\otimes\gamma^{\otimes (n_2-3)}).
  $$
Obviously,
this is a particular case of (3.3).

\smallskip

{\it Remark.} A polarization argument shows that, vice versa, (4.13)
implies all the quadratic relations (3.3).

\smallskip

We will show now that the grading conditions imply an additional
symmetry of the $(A,g)$--structure with potential $\Phi_{\omega}.$

\medskip

{\bf 4.6. Scaling.} If $\{\gamma_1,\dots \gamma_n\} =
\{\Delta_a\ \roman{with\ multiplicity\ } n_a,\ a=0,\dots ,D\},$
then the grading condition (3.2) for non-vanishing summands
of $\Phi_{\omega}$ becomes
  $$
\sum_{a=0}^Dn_a(|\Delta_a|-2)-2\int_{\beta}c_1(V)=2(\roman{dim}_{\bold{C}}V-3),
\eqno(4.20)
  $$
This leads to the following flows on our geometric data.

\medskip

{\bf 4.6.1. Scaling transformation of the potential.}
  $$ x^a\mapsto e^{(|\Delta_a|-2)t}x^a,  $$
  $$ \omega\mapsto \omega+2c_1(T(V))t,  $$
  $$ \Phi\mapsto e^{2(\dim_{\bold{C}}V-3)t}\Phi.  $$
(See (4.17).)

\medskip

{\bf 4.6.2. Scaling transformation of the Dubrovin structure.}
Since $A_{abc}$ is the tensor of the third derivatives
of $\Phi_{\omega},$ the Proposition 4.4 allows one to replace
the flow shift of $\omega$ by the reverse shift
of $\gamma^2=\sum_{|\Delta_b|=2}
x^b\Delta_b$ by $-2c_1(T(V))t.$ Defining the numbers
$\{ \xi^b\ |\ |\Delta_b|=2\}$ by
  $$
c_1(T(V))=\sum_{|\Delta_b|=2}\xi^b\Delta_b,
  $$
we obtain the following flow: $\omega \mapsto \omega ,$ and
  $$
x^a\mapsto e^{(|\Delta_a|-2)t}x^a,\qquad |\Delta_a|\ne 2;
  $$
  $$
x^b\mapsto x^b-2\xi^bt,\qquad |\Delta_b|=2;
\eqno(4.21)
  $$
  $$
A_{abc}\mapsto e^{(2\roman{dim}_{\bold{C}}V-
|\Delta_a|-|\Delta_b|-|\Delta_c|)t}A_{abc}.
\eqno(4.22)
  $$
Since the Poincar\'e pairing is invariant, and $g_{ab}\ne 0$
only for $|\Delta_a|+|\Delta_b|=2\roman{dim}_{\bold{C}}V,$
from these basic formulas we get furthermore
  $$
\partial_a\mapsto e^{(2-|\Delta_a|)t}\partial_a,\
dx^a\mapsto e^{(|\Delta_a|-2)t}dx^a;
 $$
 $$
A_{ab}^c\mapsto e^{(|\Delta_c|-|\Delta_a|-|\Delta_b|)t}A_{ab}^c,
\eqno(4.23a)
 $$
and finally, from (4.3),
 $$
\nabla_{\lambda}\mapsto \nabla_{0}+e^{2t}(\nabla_{\lambda}-\nabla_0).
\eqno(4.23b)
 $$

We now introduce an extended supermanifold $\widehat{H}^V=H^V\times\bold{P}^1$
where $\bold{P}^1$ is the completion of the affine line with coordinate
$\lambda,$ and extend the flow (4.21) to $\widehat{H}^V$ by
 $$   \lambda\mapsto e^{2t}\lambda \eqno(4.24)    $$
which is another form of (4.23b).

 From now on, $\omega$ can and will be assumed fixed.

\medskip

{\bf 4.6.3. Infinitesimal form.} The flow (4.21) can be written
as $\exp (tX)$ where $X$ is the following even vector field on
$H^V:$
 $$
X=\sum_a(|\Delta_a|-2)x^a\partial_a-2\sum_{|\Delta_b|=2}\xi^b\partial_b.
\eqno(4.25)
 $$
In view of (4.24), its natural extension to $\widehat{H}^V$ is
 $$
Y=2\lambda\frac{\partial}{\partial \lambda}+X.
\eqno(4.26)
 $$
Let $\pi:\  \widehat{H}^V\to H^V$ be the projection, and $\widehat{\Cal{T}}=
\pi^*(\Cal{T}_{H^V}).$ For a local section $\partial$ of
$\Cal{T}_{H^V},$ we denote by $\widehat{\partial}$ its lift to
$\widehat{\Cal{T}}.$ We will now extend $\nabla_{\lambda}$
to a connection on $\widehat{\Cal{T}}.$

\medskip

\proclaim{\quad 4.7. Proposition} Put
 $$
\widehat{\nabla}_{\partial_b}(\widehat{\partial }_a):=
(\nabla_{\lambda ,\partial_b}(\partial_a))\ \widehat{}=
\lambda(\partial_b\circ\partial_a)\ \widehat{}=
\lambda\sum_cA_{ba}^c\widehat{\partial }_c,
 $$
 $$
\widehat{\nabla}_Y(\widehat{\partial}_a):=(2-|\Delta_a|)\widehat{\partial}_a.
\eqno(4.27)
 $$
Then $\widehat{\nabla}$ is a flat connection on $\widehat{\Cal{T}}.$
\endproclaim

\smallskip

{\bf Proof.} Clearly, $(\partial_b,\partial_c)$--components
of the curvature vanish because of (4.7a). It remains to check that
 $$
\widehat{\nabla}_{[\partial_b,Y] }(\widehat{\partial}_a)=
[\widehat{\nabla}_{\partial_b },\widehat{\nabla}_Y](\widehat{\partial}_a).
\eqno(4.28)
 $$

We have
 $$
[\partial_b,Y]x^c=(|\Delta_c|-2)\delta_{bc},\ [\partial_b,Y]\lambda =0,
 $$
so that
 $$
[\partial_b,Y]=(|\Delta_b|-2)\partial_b,
 $$
and
 $$
\widehat{\nabla}_{[\partial_b,Y]}(\widehat{\partial}_a)=
\lambda (|\Delta_b|-2)\sum_cA_{ba}^c\widehat{\partial}_c.
\eqno(4.29)
 $$
On the other hand, in view of (4.27), (4.23),
 $$
\widehat{\nabla}_{\partial_b}\widehat{\nabla}_Y(\widehat{\partial}_a)=
\lambda (2-|\Delta_a|)\sum_cA_{ba}^c\widehat{\partial}_c,
 $$
 $$
\widehat{\nabla}_Y\widehat{\nabla}_{\partial_b}(\partial_a)=
\widehat{\nabla}_Y(\lambda\sum_cA_{ba}^c\widehat{\partial}_c)=
 $$
 $$
2\lambda \sum_cA_{ba}^c\widehat{\partial}_c+\lambda\sum_c
X(A_{ba}^c)\widehat{\partial}_c+
\lambda\sum_cA_{ba}^c(2-|\Delta_c|)\widehat{\partial}_c=
 $$
 $$
2\lambda\sum_cA_{ba}^c\widehat{\partial}_c
+\lambda\sum_c(|\Delta_c|-|\Delta_b|-|\Delta_c|)
A_{ba}^c\widehat{\partial}_c+
\lambda\sum_cA_{ba}^c(2-|\Delta_c|)\widehat{\partial}_c=
 $$
 $$
(4-|\Delta_b|-|\Delta_a|)\lambda\sum_cA_{ba}^c\widehat{\partial}_c
 $$
so that
 $$
[\widehat{\nabla}_{\partial_b},\widehat{\nabla}_Y](\widehat{\partial}_a)=
(|\Delta_b|-2)\lambda\sum_cA_{ba}^c\widehat{\partial}_c,
 $$
which agrees with (4.29).

\medskip

{\bf 4.8. Horizontal sections.} Consider an even local section
$\psi =\sum_a\psi^a\widehat{\partial}_a$ of $\widehat{\Cal{T}}.$
 From (4.27) we find
 $$
\widehat{\nabla}_{{\partial}_b}(\psi )=\sum_a
[(\partial_b\psi^a)\widehat{\partial}_a+
(-1)^{ab}\lambda\sum_c\psi^aA_{ba}^c\widehat{\partial}_c],
 $$
 $$
\widehat{\nabla}_Y(\psi )=\sum_a[Y\psi^a+
(2-|\Delta_a|)\psi^a]\widehat{\partial}_a.
 $$
Hence $\psi$ is horizontal iff
 $$
\forall b,a,\qquad \partial_b\psi^a=
-\lambda\sum_c\psi^cA_{cb}^a,
\eqno(4.30)
 $$
 $$
\forall a,\qquad Y\psi^a+(2-|\Delta_a|)\psi^a=0.
\eqno(4.31)
 $$
But
 $$
Y\psi^a=2\lambda\frac{\partial \psi^a}{\partial \lambda}+
\sum_c(|\Delta_c|-2)x^c\partial_c\psi^a-
2\sum_{|\Delta_b|=2}\xi^b\partial_b\psi^a.
 $$
Replacing here the $\partial$--derivatives of $\psi$ by the r.h.s.
of (4.30), we get finally the equation governing the
$\lambda$--dependence of the horizontal sections:
 $$
2\lambda\frac{\partial \psi^a}{\partial \lambda}
-\lambda\sum_c(|\Delta_c|-2)x^c\sum_e\psi^eA_{ec}^a+
 $$
 $$
2\lambda\sum_{|\Delta_b|=2}\xi^b
\sum_e\psi^eA_{eb}^a+(2-|\Delta_a|)\psi^a=0,
 $$
or else
 $$
2\lambda\frac{\partial\psi^a}{\partial\lambda}
-\lambda\sum_e\psi^e({\sum_c(|\Delta_c|-2)
x^cA_{ce}^a-2\sum_{|\Delta_b|=2}\xi^bA_{be}^a})
+(2-|\Delta_a|)\psi^a=0.
\eqno (4.32)
 $$
The equation (4.32) has two singular points of which
$\lambda =0$ is regular, but $\lambda =\infty$
is irregular one.

\smallskip

Therefore we make a formal Fourier transform (of $\hat{\Cal{T}}$
as a relative $\Cal{D}$--module in $\lambda$--direction)
 $$
\lambda\mapsto\frac{\partial}{\partial\mu},\
\frac{\partial}{\partial\lambda}\mapsto -\mu ,\
\psi^a\mapsto\varphi^a.
 $$
Denote the resulting $\Cal{D}$--module $\widetilde{\Cal{T}}.$

\smallskip

In order to describe it explicitly, it is convenient to pass from the
language of connections to that of quantum multiplication.

\smallskip

For any $\gamma\in H^V,\ \gamma =\sum x^c\Delta_c,$\ put
 $$
K(\gamma ):=\sum_{|\Delta_b|=2}\xi^b\Delta_b-
\sum_a (|\Delta_a|-2)x^a\Delta_a=
-K_V-\Cal{E}(\gamma)+2\gamma
\eqno (4.33)
 $$
where $\Cal{E}$ is the Euler grading operator multiplying $\Delta$
by $|\Delta |.$ Denote by $B(\gamma )$ the operator of quantum
multiplication by $K(\gamma )$ {\it at the point} $\gamma$ that is,
in $\Cal{T}_{\gamma}.$

\smallskip

We will identify $B(\gamma )$ with the matrix describing its action
from
the left to the column vector $(\Delta_a)$, and consider it also
as a matrix acting from the right on the row of coordinates of a section
$(\varphi^a)$ of $\widetilde{\Cal{T}}.$ Similar convention applies to
$\Cal{E}.$

\medskip

\proclaim{\quad 4.9. Proposition} If $B(\gamma )$ is semisimple
in a subdomain of $H^V,$
 $\widetilde{\Cal{T}}$ is an
isomonodromy deformation of a holonomic $\Cal{D}$--module
on $\bold{P}^1$ with $\le \roman{dim}\ H^*(V) +1$ regular singular
points parametrized by this subdomain.
\endproclaim

\smallskip

{\bf Proof.} The equation (4.32) becomes in $\widetilde{\Cal{T}}:$
 $$
2\mu\frac{\partial\varphi^a}{\partial\mu}+\sum_e
\frac{\partial\varphi^e}{\partial\mu}
\left[ \sum_c(|\Delta_c|-2)x^cA_{ce}^a-2\sum_{|\Delta_b|=2}
\xi^bA^a_{be}\right]
+|\Delta_a|\varphi^a=0.
 $$
On the other hand,
 $$
\Delta_e\circ_{\gamma}K(\gamma )=
-2\sum_{|\Delta_b|=2}\xi^b\sum_aA_{eb}^a\Delta_a+
\sum_c(|\Delta_c|-2)x^c(-1)^{ec}\sum_aA^a_{ec}\Delta_a.
 $$
Comparing these expressions, one sees that the Fourier transform of
(4.32) can be written in
the form
 $$
\frac{\partial\varphi}{\partial\mu}
(B(\gamma )-\mu E)=\frac{1}{2}\varphi\Cal{E}
\eqno (4.34)
 $$
where $E$ is the identity matrix.
One easily sees now that singularities of $\widetilde{\Cal{T}}$
consist of the spectrum of $B(\gamma )$ and infinity, and
that they are regular when $B(\gamma )$ is semisimple.

\smallskip

This finishes the proof.

\medskip

{\bf 4.10. Schlessinger and Painlev\'e  equations.} If we want now to
understand
the analytic properties of $\Phi$, or rather the associated
structure constants $A_{ab}^c$, as a function of $\gamma$ using (4.34),
we have to solve the following problems.

\smallskip

a). Find an appropriate point of $H^V$ where $B(\gamma )$ is
computable and semisimple.

\smallskip

We will see in a moment that for Fano varieties
 $\gamma =0$ is the first choice.

\smallskip

Denote by $M$ a realization of the moduli space of isomonodromy
deformations of the relevant $\Cal{D}$--module constructed by Malgrange.

\smallskip

b). Calculate the (partial) map $q: H^V\to M$ inducing $\widetilde{\Cal{T}}.$

\smallskip

This map can be considered as an analog, or a version, of the mirror
map discovered in the context of Calabi--Yau manifolds,
especially if the following problem can be solved affirmatively.

\smallskip

c). Identify (4.34) as an equation for the horizontal sections
of the canonical connection of a variation of Hodge structure.

\smallskip

In the classical language, the compatibility conditions
imposed upon $A_{ab}^c$ by (4.34) are called Schlessinger equations.

\smallskip

When there are only 4 singularities (the only non-trivial example being
$V=\bold{P}^2$), the isomonodromy deformations are governed
by the Painlev\'e VI equation, which is obtained from the
corresponding Schlessinger equation by a change of variables
whenever the monodromy group can be reduced to $SL(2)$.
 In this sense, the quantum cohomology
of $\bold{P}^2$ furnishes a solution of this equation, whereas
other $V$ lead to generalized Painlev\'e equations.

\medskip

{\bf 4.11. Quantum multiplication at $\gamma =0.$} From the discussion
in 4.3, 4.4, and 2.2, one sees that for Fano manifolds
the structure constants at $\gamma =0$ are defined by a finite set
of basic Gromov--Witten numbers:
 $$
A_{abc}(0)=\int_V\Delta_a\wedge\Delta_b\wedge\Delta_c+
\sum_{\beta\ne 0}e^{-\int \omega\wedge\beta}
\langle I^V_{0,3,\beta} \rangle(\Delta_a\otimes\Delta_b\otimes\Delta_c )
\eqno (4.35)
 $$
where the summation is taken over $\beta$ with
 $$
(-K_V.\beta )=\frac{1}{2}(|\Delta_a|+|\Delta_b|+|\Delta_c|)-
\roman{dim}_{\bold{C}}V\le 2\roman{dim}_{\bold{C}}V.
\eqno (4.36)
 $$
In 5.3--5.4 below, we will calculate these coefficients for
$\bold{P}^n$ and show that $B(0)$ is indeed semisimple
with simple spectrum.

\newpage


\centerline{\bf \S 5. Examples}

\bigskip

{\bf 5.1. The structure of $\Phi .$}  In this section, we
discuss in more detail some classes of manifolds $V.$
We always tacitly assume that at least tree level
GW-classes for $V$ exist, and that the potential $\Phi$
constructed from them satisfies one of the convergence
hypotheses of 4.3.

\smallskip

We start with (4.17) and take into account the Proposition 4.4
in order to drop the redundant terms, in particular those
that are no more than quadratic in $\gamma .$
Our conventions are:

\smallskip

a). $\omega =0$ (because this can be achieved by shifting $\gamma$).

\smallskip

b). $D=r+1,\ \Delta_0=e^0_V,\ \Delta_{r+1}=e_V^{2\roman{dim}_{\bold{C}}V}=$
the dual class of a point.

\smallskip

We also assume that $H^1(V)=0$ so that $2\le |\Delta_a|\le
2\roman{dim}_{\bold{C}}V-2$ for $a=1,\dots ,r.$

\smallskip

The coordinates are renamed: $\gamma =x\Delta_0+\sum_{a=1}^ry^a\Delta_a+
z\Delta_{r+1}.$

\smallskip

 From the proof of Proposition 4.4 one knows that only the
$(\beta=0,n=3)$--term
in (4.17) depends on $x$, and is
$\frac{1}{6}\int_V\gamma\wedge\gamma\wedge\gamma :=
\frac{1}{6}(\gamma^3)$ (see (2.8)). The $(\beta =0,n>3)$--terms
all vanish because of the grading condition (4.20) combined with (2.8).
So we start with
 $$
\Phi^V(\gamma )=\frac{1}{6}(\gamma^3)+
 $$
 $$
\sum_{n_1+\dots +n_{r+1}=n\ge 3}\sum_{\beta\ne 0}
\frac{\varepsilon (n_1,\dots ,n_{r+1})}{n_1!\dots n_{r+1}!}
\langle I^V_{0,n,\beta}\rangle
(\Delta_1^{\otimes n_1}\otimes\dots\otimes\Delta_{r+1}^{\otimes n_{r+1}})
(y^1)^{n_1}\dots (y^r)^{n_r}z^{n_{r+1}}
\eqno(5.1)
 $$
and the grading condition for non-vanishing terms
 $$
\sum_{a=1}^{r+1}n_a(|\Delta_a|-2)=2(-K_V.\beta)
+2(\roman{dim}_{\bold{C}}V-3).
\eqno(5.2)
 $$

\medskip

{\bf Dimension 1.} Let $V=\bold{P}^1,\ r=0,\
\beta =d[\bold{P}^1],\ d\ge 1.$ From (5.2) it follows that
$(d\ge 2)$--terms vanish. In view of the Divisor Axiom,
 $$
\langle I^{\bold{P}^1}_{0,n,[\bold{P}^1]}\rangle
(\Delta_1^{\otimes n})=
\langle I^{\bold{P}^1}_{0,3,[\bold{P}^1]}\rangle
(\Delta_1^{\otimes 3})
\eqno(5.3)
 $$
for all $n\ge 3$, and this must be 1, which is the number of
the automorphisms of $\bold{P}^1$ fixing three points.
So finally we find a (conditional) answer:

\medskip

\proclaim{\quad 5.1.1. Proposition}
 $$
\Phi^{\bold{P}^1}(xe^0+ze^2)=\frac{1}{2}x^2z+e^z-1-z-\frac{z^2}{2}.
\eqno(5.4)
 $$
\endproclaim

\smallskip

In the following calculations, we will be omitting quadratic, linear,
and constant terms of $\Phi$ without changing its notation.

\medskip

{\bf Dimension 2.} Here $\{\Delta_1,\dots ,\Delta_r\}$ form a basis
of $H^2(V);$ (5.2) is equivalent to
 $$
n_{r+1}=(-K_V.\beta)-1:=k(\beta )-1,
\eqno(5.5)
 $$
so that we must sum only over $\beta$ with $k(\beta )\ge 1.$

\smallskip

\proclaim{\quad 5.1.2. Proposition} For surfaces $V$ we have up to
terms no more than quadratic in $\gamma$
 $$
\Phi^V(\gamma )=\frac{1}{6}(\gamma^3)+
\sum_{\beta}N(\beta )\frac{z^{k(\beta )-1}}{(k(\beta )-1)!}
e^{(\beta .\gamma )}.
\eqno(5.6)
 $$
Here for $k(\beta )\ge 4$ we put
 $$
N(\beta )=\langle I^V_{0,k(\beta )-1,\beta}\rangle
(\Delta_{r+1}^{\otimes (k(\beta )-1)}).
\eqno(5.7)
 $$
For $k(\beta )\le 3,$ the definition of $N(\beta )$ is given
below: see (5.9).

\endproclaim

{\bf Proof.} If $(-K_V.\beta )\ge 4,$ then $n_{r+1}\ge 3$
(see (5.5)) so that the contribution of $\beta $ to (5.1)
in view of the Divisor Axiom takes form
 $$
\frac{z^{k(\beta )-1}}{(k(\beta )-1)!}\sum_{n_i\ge 0}
\frac{1}{n_1!\dots n_r!}\langle I^V_{0,n,\beta}\rangle
(\Delta_1^{\otimes n_1}\otimes\dots\otimes\Delta_r^{n_r}
\otimes\Delta_{r+1}^{\otimes (k(\beta )-1)}
(y^1)^{n_1}\dots (y^r)^{n_r}=
 $$
 $$
\frac{z^{k(\beta )-1}}{(k(\beta )-1)!}\sum_{n_i\ge 0}
\frac{1}{n_1!\dots n_r!}\langle I^V_{0,k(\beta )-1,\beta}\rangle
(\Delta_{r+1}^{\otimes (k(\beta )-1)})
((\beta .\Delta_1)y^1)^{n_1}\dots ((\beta .\Delta_r)y^r)^{n_r}=
 $$
 $$
N(\beta )\frac{z^{k(\beta )-1}}{(k(\beta )-1)!}e^{(\beta .\gamma )}.
\eqno(5.8)
 $$

For $1\le k(\beta )\le 3$ the calculation is only slightly longer.
The actual contribution of $\beta $ is given by the same formula
as the first expression in (5.8), but this time with summation
taken over $n_1+\dots +n_r\ge 4-k(\beta ).$ First, this sum lacks
the terms of total degree $\le 2$ in $y^i,z$, but they are negligible.
Second, in order to represent this sum as the last expression
in (5.8), we are bound to put
 $$
N(\beta ):=\frac{\langle I^V_{0,n,\beta}\rangle
(\Delta_1^{\otimes n_1}\otimes\dots\otimes\Delta_r^{\otimes n_r}
\otimes\Delta_{r+1}^{\otimes (k(\beta )-1)})}
{(\beta .\Delta_1)^{n_1}\dots (\beta .\Delta_r)^{n_r}}.
\eqno(5.9)
 $$
For fixed $n_i,$ the r.h.s. is well defined if $(\beta .\Delta_i)\ne 0$
for all $i.$ One can secure this by choosing an appropriate basis
(eventually depending on $\beta $). The result does not depend on $n_i.$
In fact, one can reach any point $(n_1,\dots n_r)$ in the set
$\sum n_i\ge 4-k(\beta ), n_i\ge 0,$ from any other point, without
ever leaving this set, by adding and subtracting 1's from  coordinates.
In view of the Divisor Axiom, these steps multiply the numerator
and the denominator of (5.9) by the same amount.

\smallskip

We expect that $N(\beta )$ counts the number of rational curves
in the homology class $\beta$ passing through $k(\beta )-1$
points, at least in unobstructed problems.

\medskip

{\bf Dimension 3.} In this dimension, Calabi--Yau manifolds make
their first appearance, and we consider their potentials.
Since $K_V=0,$ (5.2) shows that $n_a\ne 0$ only for $|\Delta_a|=2.$
Therefore, we may and will disregard the other elements of the basis
of $H^*(V),$ and in this subsection denote by $\{\Delta_1,\dots ,\Delta_r\}$
a basis of $H^2(V).$

\smallskip

\proclaim{\quad 5.1.3. Proposition} For a threefold $V$ with $K_V=0,$
we have, up to terms of degree $\le 2$ in $\gamma ,$
 $$
\Phi^V=\frac{1}{6}(\gamma^3)+\sum_{\beta\ne 0}\tilde{N}(\beta )
e^{(\beta .\gamma)}, \eqno(5.10)
 $$
where
 $$
\tilde{N}(\beta )=\frac{\langle I^V_{0,n,\beta}\rangle
(\Delta_1^{\otimes n_1}\otimes\dots\otimes\Delta_r^{\otimes n_r})}
{(\beta .\Delta_1)^{n_1}\dots (\beta .\Delta_r)^{n_r}}
\eqno(5.11)
 $$
for any $n=n_1+\dots +n_r\ge 3$ and any basis of $H^2(V)$ such that
$(\beta .\Delta_i)\ne 0$
for all $i.$

\endproclaim

\smallskip

The proof does not differ much from the previous one.

\smallskip

Let us guess the geometric meaning of $\tilde{N}(\beta )\in \bold{Q}$
 restricting
ourselves to the case $r=\roman{dim}\ H^2(V)=\roman{rk}\ Pic(V)=1.$
Let $\Delta_1 $ be the ample generator of $H^2(V),\ \beta_0$
the effective generator of $H_2(V,\bold{Z})$ with $(\beta_0 .\Delta_1)=1,$
and $\beta =d\beta_0.$ If $N(d)\in \bold{Z}$
is the ``geometric'' number of unparametrized
rational curves in the class $d\beta_0,$ then the number of
primitively parametrized
curves with three marked points landing in $\beta$ and incident
to three fixed cycles dual to $\Delta_1$ must be $d^3N(d).$

\smallskip

According to [AM], the parametrizations of degree $m$ with three marked points
 must be counted
with multiplicity $m^{-3}.$ Hence we expect that
 $$
\tilde{N}(d\beta_0)=\sum_{k/d}N(k)\frac{k^3}{d^3}
\eqno(5.12)
 $$
which can also be taken as a formal definition of numbers $N(k)$
via GW-classes. Miraculously, all
 classes appear to be integral.
 Rewriting (5.10) in this situation, we get for $y=y^1$
 $$
\Phi^V(\gamma )=\frac{1}{6}(\gamma^3)+
\sum_{k\ge 1}N(k)Li_3(e^{ky})
\eqno(5.13)
 $$
where $Li_3(z)=\sum_{m\ge 1}\frac{z^m}{m^3}.$

\medskip

{\bf Projective spaces.} Let $V=\bold{P}^{r+1},\ \Delta_i=c_1(\Cal{O}(1))^i,
|\Delta_i|=2i,\ r\ge 1.$ Put for $\sum_{a=2}^{r+1}n_a(a-1)=
(r+2)d+r-2$ (this is (5.2)),
 $$
N(d;n_2,\dots ,n_{r+1})=
\frac{\langle I^V_{0,n,d[\bold{P}^1]}\rangle
(\Delta_1^{\otimes n_1}\otimes\dots\otimes\Delta_{r+1}^{n_{r+1}})}
{d^{n_1}},
\eqno(5.14)
 $$
where $n=n_1+\dots +n_{r+1}\ge 3.$ The r.h.s. of (5.14) being independent
on $n_1$, for $n_2+\dots +n_{r+1}\ge 3$ one can take $n_1=0.$
Again, (5.14) must be the number of rational curves of degree
$d$ in $\bold{P}^{r+1}$ intersecting $n_a$ hyperplanes of codimension $a,\
a=2,\dots ,r+1.$

\smallskip

A version of previous calculations now gives:

\medskip

\proclaim{\quad 5.1.4. Proposition}
 $$
\Phi^{\bold{P}^{r+1}}(\gamma )=\frac{1}{6}(\gamma^3)+\sum_{d,n_a}
N(d;n_2,\dots ,n_{r+1})\frac{(y^2)^{n_2}\dots (y^{r+1})^{n_{r+1}}}
{n_2!\dots n_{r+1}!}e^{dy_1}.
\eqno(5.15)
 $$

\endproclaim

{\bf 5.2. Enumerative predictions.} We start with projective plane:
(5.6) for $r=1.$ Put
 $$
\varphi (y,z)=\Phi^{\bold{P}^2}(\gamma ) -\frac{1}{6}(\gamma^3)=
\sum_{d=1}^{\infty}N(d)\frac{z^{3d-1}}{(3d-1)!}e^{dy},\ N(1)=1.
 $$
We have the following simple

\medskip

\proclaim{\quad 5.2.1. Claim} The associativity condition for the
potential $\Phi^{\bold{P}^2}$ is equivalent to the single equation
 $$
\varphi_{zzz}=\varphi^2_{yyz}-\varphi_{yyy}\varphi_{yzz}
\eqno(5.16)
 $$
which in turn is equivalent to the recursive relation
 $$
N(d)=\sum_{k+l=d}N(k)N(l)k^2l\left[ l{{3d-4}\choose{3k-2}}-
k{{3d-4}\choose{3k-1}}\right] ,\ d\ge 2,
\eqno(5.17)
 $$
uniquely defining $N(d)$ and $\varphi .$

\endproclaim

\smallskip

This discovery made by M. Kontsevich was the starting point for this
paper. The first values of $N(d),$ starting with $d=2$, are
1,12,620,87304,26312976,14616808192.

\smallskip

 From 3.1 and 3.1.1 it follows that a similar uniqueness result
holds for any projective space: in the notation of (5.14), (5.15),
we have

\medskip

\proclaim{\quad 5.2.2. Claim} The associativity relations
together with the
initial condition $N(1;0,\dots ,0,2)=1$ uniquely define
all $N(d;n_2,\dots ,n_{r+1})$ and the potential $\Phi^{\bold{P}^{r+1}}.$

\endproclaim

\smallskip

Here, however, the compatibility of the associativity relations must be
established either geometrically (via a construction of GW-classes),
or by number theoretic and combinatorial means. We will now look
at some of the identities implied by associativity for del Pezzo
surfaces.

\medskip

{\bf Del Pezzo surfaces.} Let $V=V_r$ be a del Pezzo surface
that can be obtained by blowing up $0\le r\le 8$ points
(in sufficiently general position) of $\bold{P}^2$.
A choice of such a representation $\pi :V_r\to \bold{P}^2$
allows one to identify $Pic(V_r)$ with $\bold{Z}^{r+1}$
via
 $$
L=a\Lambda -b_1l_1-\dots -b_rl_r \mapsto (a,b_1,\dots b_r)
 $$
where $\Lambda =\pi^*(c_1(\Cal{O}(1)))$ and $l_i=$ inverse
image of the $i$--th blown point. Under this identification,
the intersection index becomes $((a,b_i).(a^{\prime},b_i^{\prime}))
=aa^{\prime}-\sum b_ib_i^{\prime},$ and $-K_V=(3;1,\dots ,1)$
so that $(-K_V.L)=3a-\sum b_i.$ The cone of effective classes
$B$ is generated by its indecomposable elements
$\Lambda$ for $r=0,$\ $\Lambda -l_1$
and $l_1$ for $r=1,$ and all exceptional classes for $r\ge 2.$
(Recall that $l$ is exceptional iff $(l^2)=-1$ and $(-K_V.\beta )=1;$
for more details see [Ma1]).

\smallskip

This allows us to rewrite (5.6) as an explicit sum over $B.$

\smallskip

Writing $\gamma =xe^0+y\Lambda +ze^4-\sum_{i=1}^ry^il_i,$
$\Phi^{V_r}(\gamma )=\frac{1}{6}(\gamma^3)+\varphi (\gamma),$
we can easily check the following generalization of  5.2.1:

\medskip

\proclaim{\quad 5.2.3. Claim} a). One of the associativity relations
reads
 $$
\varphi_{zzz}=\varphi^2_{yyz}-\sum_{i=1}^r \varphi^2_{yy^iz}-
\varphi_{yyy}\varphi_{yzz}+\sum_{i=1}^r\varphi_{yyy^i}\varphi_{y^izz}.
 $$

b). This relation is equivalent to the following recursive formula for
the coefficients $N(\beta )$ (see (5.6)):
 $$
N(\beta )=
 $$
 $$
\sum_{\beta_1+\beta_2=\beta}N(\beta_1)N(\beta_2)(\beta_1.\beta_2)
(\Lambda .\beta_1)\left[ (\Lambda .\beta_2){{(-K_V.\beta )-4}
\choose{(-K_V.\beta_1)-2}}
-(\Lambda .\beta_1){{(-K_V.\beta )-4}\choose{(-K_V.\beta_1)-1}}\right] .
\eqno(5.18)
$$

\endproclaim

\smallskip

The initial conditions for (5.18) consist of the list of values of $N(\beta )$
for all indecomposable elements of $B.$ It is expected that all these
values are 1.

\smallskip

The redundancy of the associativity relations is reflected here
in the presence of $\Lambda$ which depends on the choice
of $\pi :V_r\to\bold{P}^2.$ The number $c_r$ of such choices
for $r=1,\dots ,8$ is respectively $1,1,2,5,2^4,2^3.3^2,2^6.3^2,
2^7.3^3.5.$ In fact, the symmetry group $W_r$ of the configuration
of exceptional classes acts upon the set of associativity relations, and
$c_r=|W_r|/r!$, the denominator corresponding to the renumberings
of blown points.

\smallskip

{\it Question.} Is it true that the linear span of all relations
(5.18) for various choices of $\pi$ contains all the associativity
relations, at least for larger values of $r$?

\medskip

{\bf 5.2.4. Quadric.} The quadric $V=\bold{P}^1\times\bold{P}^1$
is the last del Pezzo surface. Here $Pic(V)=\bold{Z}^2,\ -K_V=(2,2),$
and all associativity relations were written explicitly in [I].
In self-explanatory notation
 $$
\varphi (\gamma )=\sum_{a+b\ge 1}N(a,b)\frac{z^{2a+2b-1}}{(2a+2b-1)!}
e^{ay^1+by^2},
 $$
and the associativity relations together with initial conditions
$N(0,1)=N(1,0)=1$
imply the following recursive definition of $N(a,b)$ in the
effective cone $a\ge 0,b\ge 0:$
 $$
N(a,b)=
 $$
 $$
\sum_{{a_1+a_2=a}\atop{b_1+b_2=b}}N(a_1,b_1)N(a_2,b_2)
(a_1b_2+a_2b_1)b_2\left[ a_1{{2a+2b-4}\choose{2a_1+2b_1-2}}-
a_2{{2a+2b-4}\choose{2a_1+2b_1-1}}\right] .
\eqno(5.19)
$$
The remaining relations are:
 $$
N(a,b)=N(b,a), \eqno(5.20)
 $$
 $$
2abN(a,b)=\sum_{{a_1+a_2=a}\atop{b_1+b_2=b}}N(a_1,b_1)N(a_2,b_2)
{{2a+2b-3}\choose{2a_1+2b_1-1}}a_1^2(a_1b_2-a_2b_1)b_2^2,
\eqno(5.21)
 $$
 $$
aN(a,b)=\sum_{{a_1+a_2=a}\atop{b_1+b_2=b}}N(a_1,b_1)N(a_2,b_2)
{{2a+2b-3}\choose{2a_1+2b_1-1}}a_1(a_1^2b_2^2-b_1^2a_2^2),
\eqno(5.22)
 $$
 $$
0=\sum_{{a_1+a_2=a}\atop{b_1+b_2=b}}N(a_1,b_1)N(a_2,b_2)
{{2a+2b-3}\choose{2a_1+2b_1-1}}\times
 $$
 $$
a_1^2[(a_2+b_2-1)(b_1a_2+a_1b_2
)-(2a_1+2b_1-1)a_2b_2].
\eqno(5.23)
 $$
\smallskip

{\it Question.} Can one deduce (5.20)--(5.23) directly from (5.19)?

\medskip

{\bf 5.2.6. Nonsingular rational curves.} Consider an effective class
$\beta$ with \linebreak $p_a(\beta ):=(\beta .\beta +K_V)/2+1=0,$ i.e.
 $$
(-K_V.\beta )=d\ge 0,\ (\beta .\beta )=d-2.
 $$
Any irreducible curve in this class must be nonsingular rational,
so that passing through points imposes only linear conditions.
Thus we may expect that $N(\beta )=1$ for such a class.
This was observed numerically on cubic surfaces $V_6$ by
C. Itzykson.

\smallskip

{\it Question.} Can one deduce directly from (5.19) that
$N(\beta )=1$ whenever $p_a(\beta )=0$?

\smallskip

Notice that there are infinitely many such classes on any $V_r$
with $r\ge 1.$ The simplest family is: $r=1,\beta =n\Lambda -
(n-1)l_1$ projecting into rational curves of degree $n$
with one $(n-1)$--ple point on $\bold{P}^2.$

\medskip

{\bf 5.3. Quantum multiplication in $H^*(\bold{P}^n)$ at $\gamma =0.$}
Choose as above $\Delta_a=c_1(\Cal{O}(1))^a,\ |\Delta_a|=2a,\
0\le a \le n.$ Calculate $A_{abc}$ with the help of (4.35).
(We now drop the restriction $\omega =0.$ Equivalently, we can say that
we calculate the quantum multiplication with $\omega =0$ but
at all points of the subspace $H^2\subset H^*$). The $\beta \ne 0$
terms in (4.35) do not vanish only for $\beta =$ class of a line,
$1\le a,b,c \le n,\ a+b+c=n+1.$ Put $q=e^{-\int\omega\wedge\beta}.$
The compatibility of WDVV--equations implies that
$\langle I_{0,3,\beta}\rangle (\Delta_a\otimes\Delta_b\otimes\Delta_c)=1$
in this range; this agrees also with geometric interpretation.
Finally, $g_{ab}=\delta_{a+b,n}.$ Putting this all together,
we obtain:

\proclaim{\quad 5.3.1. Proposition} $(\Delta_1)^{\circ a}= \Delta_a$
for $0\le a \le n,$ and $q\Delta_0$ for $a=n+1.$
\endproclaim

Since $\Delta_0$ is the identity with respect to quantum multiplication,
we see that
 $$
H^*_{quant}(\bold{P}^n)|_{\gamma =0}\cong
\bold{C}[x]/(x^{n+1}-q).
\eqno (5.24)
 $$
This formula was heuristically obtained for $\bold{P}^n$ in many
papers, and was generalized by Batyrev for toric varieties,
and by Givental and Kim for flag spaces.

\smallskip

We now see however, that (5.24) and these generalizations describe only
a subspace of quantum deformations parametrized by $H^2.$

\medskip

{\bf 5.4. $\Cal{D}$--module $\widetilde{\Cal{T}}$ at $\gamma =0.$}
We can now easily write for $V=\bold{P}^n$ the equation (4.34) at
$\gamma =0.$ In fact, the matrix $B(0)$ describes the quantum
multiplication by $-K_V=(n+1)\Delta_1$ in the basis $(\Delta_a).$
Therefore (4.34) reads:
 $$
\pmatrix
-\mu & 0 & 0 & \dots & 0 & (n+1)q \\
n+1 & -\mu & 0 & \dots & 0 & 0 \\
\vdots & \vdots & \vdots & \ddots & \vdots & \vdots \\
0 & 0 & 0 & \dots & n+1 & -\mu
\endpmatrix
\pmatrix
\partial\varphi^0/\partial\mu\\
\partial\varphi^1/\partial\mu\\
\vdots\\
\partial\varphi^n/\partial\mu
\endpmatrix
=
\pmatrix
0\\
\varphi^1\\
\vdots \\
n\varphi^n
\endpmatrix.
 $$
The finite singular points are $(n+1)q^{1/n}.$

\newpage


\centerline{\bf \S 6. Cohomological Field Theory}

\bigskip

{\bf 6.1. Definition.} A two--dimensional
cohomological field theory (CohFT)
with coefficient field $K$ consists of
the following data:

\smallskip

a). A $K$--linear superspace (of fields) $A,$ endowed with an
even non-degenerate pairing.

\smallskip

b). A family of even linear maps (correlators)
 $$
I_{g,n}:\ A^{\otimes n}\to H^*({\overline{M}}_{g,n},K),
 $$
defined for all $g\ge 0$ and $n+2g-3\ge 0.$

\smallskip

These data must satisfy the following axioms:

\smallskip

{\bf 6.1.1. $S_n$--covariance.}

\smallskip

{\bf 6.1.2. Splitting.} In order to state this axiom, we retain all
notation of 2.2.6 with the following minimal changes: $\{ \Delta_a\}$
denotes a basis of $A,\ \Delta =\sum g^{ab}\Delta_a\otimes\Delta_b$
is the Casimir element of the pairing, and all mention of $V$ and
$\beta$'s is omitted. Then the axiom reads:
 $$
\varphi^*_S(I_{g,n}(\gamma_1\otimes\cdots\otimes\gamma_n))=
 $$
 $$
\varepsilon (S)\sum_{a,b}
I_{g_1,n_1+1}(\otimes_{j\in S_1}\gamma_j\otimes\Delta_a)g^{ab}\otimes
I_{g_2,n_2+1}(\Delta_b\otimes (\otimes_{j\in S_2}\gamma_j)).
\eqno (6.1)
 $$

\smallskip

{\bf 6.1.3. Genus reduction.} This is (2.12), with $V$ and $\beta$
omitted.

\medskip

{\bf 6.2. Remarks.} a). We will mostly assume $A$ finite--dimensional.
However, $A$ can also be graded with finite-dimensional components,
or an object of a $K$--linear tensor category, etc.

\smallskip

b). We will be mostly concerned with {\it tree level} CohFT.
Correlators for such a theory must be given only for $g=0,\ n\ge 3,$
and the Genus Reduction Axiom is irrelevant.

\medskip

{\bf 6.3. Example: GW-theories.} Any system of GW-classes
for a manifold $V$ satisfying appropriate convergence
assumptions gives rise to the following cohomological field theory
depending on a class $\omega$ as in 4.3:
$A=H^*(V,\bold{C})$ with Poincar\'e pairing,
 $$
I_{g,n}:=\sum_{\beta}I^V_{g,n,\beta}e^{-\int_{\beta}\omega}.
\eqno (6.2)
 $$
The series in $\beta$ can also be treated as a formal one.
Equivalently, we can put $A=\oplus_{\beta\in B}H^*(V,K)$ and work with
$B$--graded objects.

\medskip

{\bf 6.4. Operations on field theories.} a). Tensor product of
two cohomological field theories can be defined as in 2.5.

\smallskip

b). If a group $G$ acts upon the space of fields $A$ of a tree level CohFT
preserving scalar product and $I_{g,n},$ a new tree level CohFT can be obtained
by restricting all maps to the space of invariants $A^G$.

\smallskip

c). Following Witten ([W]), one can define an infinite--dimensional
family of deformations of any CohFT. The parameter space
will be the formal neighborhood of zero in the vector
superspace $A\otimes\bold{C}[[x]]$ where $x$ is an even formal
variable. The space of fields $A$ and its scalar product
are kept undeformed.

For a point $\alpha =\sum_0^{\infty}\alpha_ix^i,\ \alpha_i\in A,$
of the parameter space, define the new correlators by
 $$
\widetilde{I}_{g,n}(\gamma_1\otimes\cdots\otimes\gamma_n)=
 $$
 $$
\sum_{k=0}^{\infty}\frac{1}{k!}\sum_{i_1,\dots ,i_k}
\pi_{k*}\left( I_{g,n+k}(\gamma_1\otimes\cdots\otimes\gamma_n\otimes
\alpha_{i_1}\otimes\cdots\otimes\alpha_{i_k})\wedge
(\wedge_{j=1}^kc_1(\Cal{T}^*_{x_{n+j}}C)^{i_j})\right) .
 $$

Here $x_i$ denotes the $i$--th marked point of the universal curve $C$,
$\pi_k$ is the projection ${\overline{M}}_{g,n+k}\to{\overline{M}}_{g,n}$
forgetting the last $k$ points, $\pi_*$ is the direct image
in cohomology for the proper map $\pi$ of smooth stacks
(orbifolds).

\medskip

{\bf 6.5. Mirror Symmetry.} Physicists believe that to each
Calabi--Yau manifold $V$ of sufficiently big K\"ahler volume
one can associate {\it two} different cohomological field
theories, called A-- and B-- models in [W].

\smallskip

A--model depends only on the cohomology class $[\omega]$ of the symplectic
form and remains invariant when one deforms the complex structure
of $V.$

\smallskip

In the infinite volume limit A-model should approximate
the GW-model from 6.3, taking its existence for granted.
One expects (see [BCOV]) that the difference
between any correlators in A-- and GW--models is a cohomology
class of a moduli space Poincar\'e dual to a homology class
supported on the boundary.

\smallskip

B--model should depend only on the complex structure of $V$
via the universal infinitesimal variation of its Hodge structure
and, possibly, on some additional data. The space of fields
in the B--model must be $\oplus_{p,q}H^q(\wedge^p\Cal{T}_V)$
with grading $(p+q)\ \roman{mod}\ 2.$ The quantum multiplication
in the computed examples is given by the symbol of iterated
canonical connection on the cohomology space whose
definition requires a choice of the global volume form on $V.$

\smallskip

Mirror symmetry ought interchange A-- and B-- models of dual
varieties.

\smallskip

It would be important to have a treatment of A-- and B-- models
axiomatizing their dependence on the geometry of $V.$

\smallskip

In the remaining part of this section, we introduce an operadic firmalism
for description of tree level CohFT's. Our framework is similar to that of
[V], [BG], [GiK].

\medskip

{\bf 6.6. Trees.} We will formally introduce trees describing combinatorial
structure of a marked stable curve of arithmetical genus $0.$
Their vertices correspond to components, and edges to special points.

\smallskip

\proclaim{\quad 6.6.1. Definition} A (stable) tree $\tau$ is a
collection of finite sets $V_\tau$ (vertices), $E_\tau$ (interior
edges), $T_\tau$ (exterior edges, or tails), and two boundary
maps $b:\ T_\tau \to V_\tau$ (every tail has one end vertex),
and $b:\ E_\tau \to  \{$unordered pairs of distinct vertices$\}$
(every interior edge has exactly two vertices).

\smallskip

The geometric realization of $\tau$ must be connected and
simply connected. Every vertex must belong to at least three
edges, exterior and/or interior (stability).

\endproclaim

\medskip

\proclaim{\quad 6.6.2. Definition} A morphism of trees $f:\ \tau
\to \sigma$ is a collection of three maps (notice arrow directions)
 $$
f_v:V_\tau \to V_\sigma ,\ f^t:\ T_\sigma \to T_\tau,\
f^e:\ E_\sigma \to E_\tau ,
 $$
with the following properties:

a). $f_v$ is surjective, $f^t$ and $f^e$ are injective.

b). If $v_1,v_2$ are ends of an edge $e^{\prime}$ of $\tau$, then
either $f_v(v_1)=f_v(v_2),$ or $f_v(v_i)$ are ends of an edge
$e^{\prime\prime}$ of $\sigma$; we say that $e^{\prime}$
covers this edge, and we must then have $e^{\prime}=f^e(e^{\prime\prime}).$

c). If $v^{\prime}\in V_\tau$ is such a vertex that $f_v(v^{\prime})$
is the end of $t^{\prime\prime}\in T_{\tau},$ then $v^{\prime}$
is the end of $f^t(t^{\prime\prime}).$

\endproclaim

\smallskip

In other words, $f$ contracts interior edges from
$E_{\tau}\setminus f^e(E_{\sigma})$
and tails from $T_{\tau}\setminus f^t(T_{\sigma})$,
and is one--to--one on the remaining edges. We will denote
by $f(e)$ the image of a non--contracted edge.

\smallskip

The composition of morphisms is the composition of maps.
In this way, trees form a category.

\medskip

{\bf 6.6.3. Flags and dimension.} A pair $\{${\it edge, one end of it}$\}$
is called a flag. For a tree $\tau ,$ we denote by $F_{\tau}$
the set of its flags, and by $F_{\tau}(v)$ the set of flags
ending in vertex $v.$ We have $|F_{\tau}|=2|E_{\tau}|+|T_{\tau}|.$

The {\it dimension} of $\tau$ is defined by
 $$
\roman{dim}\ \tau :=\sum_{v\in V_{\tau}}(|F_{\tau}(v)|-3)=
2|E_{\tau}|+|T_{\tau}|-3|V_{\tau}|.
\eqno (6.3)
 $$

\medskip

{\bf 6.6.4. Glueing.} Let $(\tau_i,t_i),\ i=1,2,$ be two pairs
consisting each of a tree and its tail. Their glueing
($t_1$ to $t_2$) produces a pair $(\tau ,e)$ consisting of a tree
and its interior edge:
 $$
(\tau ,e):=(\tau_1,t_1)*(\tau_2,t_2).
 $$
Formally:
 $$
V_{\tau}=V_{\tau_1}\coprod V_{\tau_2},\
E_{\tau}=E_{\tau_1}\coprod E_{\tau_2}\coprod \{ e\},
 $$
 $$
T_{\tau}=(T_{\tau_1}\coprod T_{\tau_2})\setminus \{ t_1,t_2 \} ,\
b(e)=\{ b(t_1),b(t_2)\} .
 $$
This operation is functorial in the following sense: for
two morphisms $f_i:\ \tau_i \to \sigma_i$ not contracting
$t_i$, we have a selfexplanatory morphism
 $$
f_1*f_2:\ (\tau_1,t_1)*(\tau_2,t_2)\to
(\sigma_1,f_1(t_1))*(\sigma_2,f_2(t_2)).
 $$
Finally, $F_{\tau}=F_{\tau_1}\coprod F_{\tau_2}.$

\medskip

{\bf 6.7. Products of families.} Assume that we work in a monoidal
category where products of families of objects are associative
and commutative in a functorial way. Then we can use
a convenient formalism (spelled out e.g., by Deligne) of
products of families indexed by {\it finite sets} and functorial
with respect to the maps of such sets. In this sense, we
will use notation like $\prod_{i\in F}A_i,\ \otimes_{i\in F}A_i,\
A^{\otimes F}$ (for a constant family $A_i=A$), etc.

\smallskip

In the same vein, if $|F|\ge 3,$ we will denote by $\overline{M}_{0,F}$
the moduli space of stable curves of genus zero with $|F|$
marked points indexed by $F.$

\medskip

{\bf 6.8. From trees to moduli spaces.} In this subsection, we
define a functor
 $$
\Cal{M}:\ \{ trees\}\to \{ algebraic\ manifolds\}
 $$
(ground field is arbitrary).

\smallskip

{\bf 6.8.1. Objects.} Put
 $$
\Cal{M}(\tau )=\prod_{v\in V_{\tau}}
\overline{M}_{0,F_{\tau}(v)}.
\eqno (6.4)
 $$
We have $\roman{dim}\ \Cal{M}(\tau )=\roman{dim}\ \tau .$

\smallskip

This space parametrizes a family of (generally reducible)
stable rational curves $C(\tau )$ with marked points
indexed by $T_{\tau }.$ The dual graph of a generic
(but not arbitrary) curve of this family is
(canonically  identified with) $\tau$. To describe it,
consider a point $x=(x_v)\in \Cal{M}(\tau ),\
x_v\in \overline{M}_{0,F_{\tau}(v)},$ and let $C(x_v)$
be the fiber of a universal curve at this point.
If $v_1,v_2$ bound an edge $e$ of $\tau ,\ C(x_v)$
contains a point $y(v_i,e)$ marked by the flag $(v_i,e).$
Identify $y(v_1,e)$ with $y(v_2,e)$ in the disjoint union
$\coprod_{v\in F_{\tau}}C(x_v)$ for all $e.$ This will be
$C(\tau )(x).$

\smallskip

Clearly, its remaining special points are marked by $T_{\tau}$
so that we have a canonical morphism (closed embedding)
$\Cal{M}(\tau )\to \overline{M}_{0,T_{\tau}}.$ This is
a special case of morphisms defined below.

\medskip

{\bf 6.8.2. Morphisms.} Any morphism of trees $f:\ \tau \to \sigma$
contracting no tails induces a closed embedding
$\Cal{M}(\tau )\to \Cal{M}(\sigma ).$ To construct it,
identify $T_{\tau}=T_{\sigma}=T$ by means of $f^t$,
and denote by $\rho$ the one--vertex tree with tails $T.$
Clearly, $\Cal{M}(\rho )=\overline{M}_{0,T},$
and by universality,
we have embeddings of $\Cal{M}(\sigma )$ and $\Cal{M}(\tau )$
into $\Cal{M}(\rho ).$ In this embedding, $\Cal{M}(\sigma )
\subset \Cal{M}(\tau )$ which is the seeked for morphism.

\smallskip

Any morphism of one--vertex trees contracting tails induces
the forgetful morphism of the respective moduli spaces
(see e.g. [Ke]).

\smallskip

The general construction of a moduli space morphism corresponding
to a morphism of trees can be obtained by combining these
two cases: embed $\Cal{M}(\tau )$ into $\overline{M}_{0,T_{\tau}},$
$\Cal{M}(\sigma )$ into $\overline{M}_{0,T_{\sigma}}$,
and restrict the forgetful map onto $\Cal{M}(\tau ).$

\medskip

{\bf 6.8.3. Glueing.} If $(\tau ,e)=(\tau_1,t_1)*(\tau_2,t_2),$
we have canonically ($H^*\Cal{M}(\cdot ):=H^*(\Cal{M}(\cdot ),K$)):
 $$
\Cal{M}(\tau )=\Cal{M}(\tau_1)\times\Cal{M}(\tau_2),
 $$
 $$
H^*\Cal{M}(\tau )=H^*\Cal{M}(\tau_1)\otimes H^*\Cal{M}(\tau_2).
\eqno (6.5)
 $$

\medskip

{\bf 6.9. From trees to tensors.} Let $A$ be a linear superspace
over $K$ with a Casimir element $\Delta $, as in 6.1.

\smallskip

{\bf 6.9.1. Objects.} For a tree $\tau$, put
$\Cal{A}(\tau )=A^{\otimes F_{\tau}}.$
We will show that this construction is (contravariant) functorial
with respect to {\it pure contractions} that is,
morphisms of trees contracting no tails.

\smallskip

{\bf 6.9.2. Morphisms.} Let $f:\ \tau\to\sigma$ be a pure
contraction, $F_{\tau}^s$ the set of non--contracted flags,
and $E_{\tau}^c$ the set of contracted edges. Each such edge
gives rise to a pair of contracted flags. Therefore, we have
two identifications:
 $$
f^{-1}:\ \Cal{A}(\sigma )=A^{\otimes F_{\sigma}}\to
A^{\otimes F_{\tau}^s},
 $$
 $$
A^{\otimes F^s_{\tau}}\otimes (A\otimes A)^{E^c_{\tau}}\to
A^{\otimes F_{\tau}},
 $$
the second being defined only up to switches in $A\otimes A$ factors.
Since the Casimir element is invariant with respect
to the switch, we can unambiguously set
 $$
\Cal{A}(f)(\otimes_{i\in F_{\sigma}}\gamma_i)=
\left( \otimes_{f^{-1}(i)\in F^s_{\tau}}
\gamma_{f^{-1}(i)})\right) \otimes \Delta^{\otimes E^c_{\tau}}.
\eqno (6.6)
 $$

\medskip

{\bf 6.9.3. Glueing.} Obviously, for $(\tau ,e)$ as in 6.8.3,
we have canonically
 $$
\Cal{A}(\tau )=\Cal{A}(\tau_1)\otimes \Cal{A}(\tau_2).
\eqno (6.7)
 $$

We see that with respect to pure contractions and glueing,
$\Cal{A}$ has the same properties as $H^*\Cal{M}$ from 6.8.

\smallskip

We can now state our new definition.

\medskip

\proclaim{\quad 6.10. Definition} An operadic tree level CohFT
is a morphism of functors compatible with glueing
 $$
I:\ \Cal{A}\to H^*\Cal{M} .
 $$
In other words, it consists of a family of maps indexed by trees
 $$
I(\tau ):\ {A}^{\otimes F_{\tau}}\to H^*(\Cal{M}(\tau ),K)
 $$
such that for any pure contraction $f:\ \tau\to\sigma$ we have:
 $$
I(\tau )\circ \Cal{A}(f)=H^*\Cal{M}(f)\circ I(\sigma ):\
{A}^{\otimes F_{\sigma}}\to H^*\Cal{M}(\tau ),
\eqno (6.8)
 $$
and for any $\tau$ glued from $\tau_1,\tau_2$,
 $$
I(\tau )=I(\tau_1 )\otimes I(\tau_2):
\Cal{A}(\tau )=\Cal{A}(\tau_1)\otimes\Cal{A}(\tau_2)\to
H^*\Cal{M}(\tau_1)\otimes H^*\Cal{M}(\tau_2).
\eqno (6.9)
 $$
\endproclaim

\medskip

\proclaim{\quad 6.11. Claim} The two definitions of a tree level
CohFT are equivalent.
\endproclaim

\smallskip

{\bf Proof (sketch).} Notice that a two--vertex tree with tails
$\{ 1,\dots ,n\}$ is the same as a partition $S=(S_1,S_2)$
of $\{ 1,\dots ,n\} .$ Now, given an operadic tree level CohFT,
restrict it to the following subclass of trees: $T_{\tau}=
\{ 1,\dots ,n\}$ for some $n,$ $|V_{\tau}|=1$ or $2.$
One easily checks that morphisms $\varphi_S$ from 2.2.6
are induced by non-trivial pure contractions in this subclass,
and that (6.7) restricted to it becomes (6.1), and
$S_n$--covariance corresponds to the functorality with respect
to bijective maps of tails.

\smallskip

Conversely, given a tree level CohFT in the sense of 6.1,
we first rewrite it as a fragment of an operadic CohFT as above,
and then reconstruct the whole operadic CohFT using glueing
and decomposition of pure contractions into products
of morphisms contracting exactly one edge each.

\newpage

\centerline{\S 7. Homology of moduli spaces}

\bigpagebreak

{\bf 7.1. Additive generators.} If $T_{\tau}=\{ 1,\dots ,n\} ,$ we will call
$\tau$ an $n$--tree. A morphism of $n$--trees $\tau \to \sigma$
identical on tails
will be called $n$--morphism. If such a morphism exists, it is unique.
Let $\rho_n$ be a one--vertex $n$--tree.
Then $\Cal{M}(\rho_n)=\overline{M}_{0,n}.$ For any $n$--tree $\tau ,$
there exists a unique $n$--contraction $\tau\to\rho_n.$ Let
$d_{\tau}\in H_*(\overline{M}_{0,n})$ be the homology class of
$\Cal{M}(\tau )$ corresponding to this contraction. It depends
only on the $n$--isomorphism class of $\tau $. The manifolds
$\Cal{M}(\tau )$ embedded into each other in this way will be called
strata.

\medskip

\proclaim{\quad Lemma 7.1.1} $d_{\tau}$ span $H_*(\overline{M}_{0,n})$
(over any coefficient ring).
\endproclaim

\smallskip

{\bf Proof.} Easy induction by $n$, as in the proof of 2.5.2.

\medskip

{\bf 7.2. Linear relations.} Choose a system $R=(\tau ,\{ i,j,k,l\} ,v),$
where $\tau$ is an $n$--tree, $1\le i,j,k,l\le n$ are its pairwise
distinct tails, and $v\in V_{\tau}$ is such a vertex that paths
from $v$ to $i,j,k,l$ start with pairwise distinct edges
$e_i,e_j,e_k,e_l$ respectively (some of these edges may be tails
themselves).

\smallskip

Consider all $n$--contractions $\tau^\prime\to\tau$ which contract
exactly one edge onto the vertex $v$ and satisfy the following
condition: lifts to $\tau^\prime$ of $e_i,e_j$ on the one hand,
and $e_k,e_l$ on the other, are incident to different vertices
of the contracted edge. Below we will denote by
$\{ ij\tau^\prime kl\}$ the summation over $n$--isomorphism classes
of such contractions, $R$ being fixed.

\smallskip

\proclaim{\quad 7.2.1. Lemma} For any $R$, we have
 $$
\sum_{\{ij\tau^\prime kl\}}d_{\tau^\prime}=
\sum_{\{ik\tau^{\prime\prime}jl\}}d_{\tau^{\prime\prime}}
\eqno (7.1)_R
 $$
in $H_*(\overline{M}_{0,n}).$
\endproclaim

\smallskip

{\bf Proof.} Consider a morphism of $\tau$ contracting all edges
and tails except of $i,j,k,l.$ It induces the forgetful
morphism $\Cal{M}(\tau )\to \overline{M}_{0,\{ ijkl\}}\cong \bold{P}^1.$
Two fibers over boundary divisors of the latter moduli
space are represented by the cycles
$\sum_{\{ ij\tau^{\prime}kl\}}\Cal{M}(\tau^{\prime})$ and
$\sum_{\{ik\tau^{\prime\prime}jl\}}\Cal{M}(\tau^{\prime\prime})$
respectively.

\medskip

\proclaim{\quad 7.3. Theorem} Relations $(7.1)_R$ span the space
of all linear relations between $d_{\tau}.$
\endproclaim

\smallskip

\proclaim{\quad 7.3.1. Lemma ([Ke])} As an algebra,
$H^*:=H^*(\overline{M}_{0,n})$ is generated by the boundary divisorial
cohomology classes $D_S$ indexed by unordered partitions
$S$ of $\{ 1,\dots ,n\}$ into two parts $S_1,S_2$ of
cardinality $\ge 2$ and satisfying the following generating relations:
 $$
\sum_{\{ ijSkl\}}D_S=\sum_{\{ ikTjl\}}D_T
\eqno (7.2)_{ijkl}
 $$
and
 $$
D_SD_T=0 \eqno (7.3)
 $$
if four sets $S_i\cap T_j$ are pairwise distinct and non--empty.
(In this case we will call $S$ and $T$ incompatible).
\endproclaim

\smallskip

In 3.2.2, $D_S$ were denoted $d_S$ whereas here we reserve
lower case letters for homology classes. Classes $D_S$ are dual
to the homology classes $d_{\sigma}$ where $\sigma$ run over
$n$--trees with two vertices, and (7.2) is a consequence of (7.1).

\smallskip

Denote now by $H_*$ the linear space generated by the symbols
$[d_{\sigma}]$ subject to all relations $(7.1)_R$ where
$\sigma ,\tau$ run over all $n$--isomorphism classes of
$n$--trees.

\smallskip

There is an obvious surjective map $a:\ H_*\to H^*:$
 $$
a([d_{\sigma}]):=\ the\ cohomology\ class\ dual\ to\ d_{\sigma} .
\eqno (7.4)
 $$

\smallskip

\proclaim{\quad 7.3.2. Main Lemma} $H_*$ can be endowed with a structure
of cyclic $H^*$--module generated by $[d_{\rho_n}]:=1$
so that the map
 $$
b:\ H^*\to H_*,\ b(h)=h\cdot 1
\eqno (7.5)
 $$
is surjective.
\endproclaim

\smallskip

Comparing (7.4) and (7.5) we see that $\roman{dim}\ H_*=\roman{dim} H^*$
and both $a$ and $b$ are linear isomorphisms. Theorem 7.3 follows.

\smallskip

{\bf Proof of 7.3.2. (Sketch).} Every interior edge $e$ of an $n$--tree
$\sigma$ determines a partition $S(\tau , e)$ of its tails.
Put
 $$
\Cal{S}(\tau )=\{ S\ |\ \exists e, S=S(\tau ,e)\} .
 $$
The stratum $\Cal{M}(\tau )$ is the transversal intersection
of pairwise compatible boundary divisors in $\Cal{M}(\rho_n):$
 $$
\Cal{M}(\tau )=\cap_{S\in \Cal{S}(\tau )}\Cal{M}(\tau_S).
 $$
Therefore we devise the action of $H^*$ upon $H_*$ in such a way that
 $$
\prod_{S\in \Cal{S}(\tau )}D_S.1=[d_{\tau}] \eqno (7.6)
 $$
making obvious the surjectivity of (7.5). If we now want to
define any product $D_S.[d_{\tau}],$ (7.6) forces us to consider
three cases.

\smallskip

{\it Case 1.} $S$ is incompatible with some $T\in \Cal{S}(\sigma ).$
Then we put
 $$
D_S.[d_{\sigma}]=0. \eqno (7.7)
 $$

\smallskip

{\it Case 2.} $S$ is compatible with all $T\in \Cal{S}(\sigma )$
but $S\notin \Cal{S}(\sigma ).$ Then we check that {\linebreak}
$\Cal{S}(\sigma )
\cup \{ S\} =\Cal{S}(\tau )$ for a unique $\tau$, and put
 $$
D_S.[d_{\sigma}]=[d_{\tau}]. \eqno (7.8)
 $$

\smallskip

{\it Case 3.} $S\in \Cal{S}(\tau ).$ This case concentrates all the
difficulties. Let us start with a two--vertex $\sigma,\ \Cal{S}(\sigma )=
\{ S\} .$ Choose $i,j\in S_1,\ k,l\in S_2$ and apply $D_S$ to the
relation between homology classes dual to $(7.2)_{ijkl}.$ We are
forced to put
 $$
D_S.[d_{\sigma}]=-\sum_{\tau}[d_{\tau}] \eqno (7.9)
 $$
where $\tau $ runs over all trees with $\Cal{S}(\tau )=\{ S,S^{\prime}\} ,$
$S^{\prime}\ne S,\ i,j\in S^{\prime}_1,\ k,l\in S^{\prime}_2,$
and then to check that, modulo postulated relations, the r.h.s.
of (7.8) does not depend on the choice of $i,j,k,l.$

\smallskip

Now, consider the r.h.s. of (7.9) as a divisorial cohomology class
in $H^*(\Cal{M}(\sigma ))\cong H^*(\overline{M}_{0,S_1})\otimes
H^*(\overline{M}_{0,S_2}).$ It can be naturally represented
as a sum of two components $D_1\otimes 1+1\otimes D_2.$
If $d_{\tau}$ is represented by $\Cal{M}(\tau )\subset \Cal{M}(\sigma ),$
we have $(\tau ,e)=(\tau_1,t_1)*(\tau_2,t_2),$ where $e$ is the lift
of the unique edge of $\sigma$ to $\tau .$ The action of $D_i$ upon
$d_{\tau_i}$ is assumed to be inductively defined, which determines
the action of $D_S$ upon $d_{\tau}.$

\smallskip

It remains to check that these prescriptions are compatible
with (7.1)--(7.3).

\smallskip

This is a tedious but straightforward verification which we omit.

\newpage


\centerline{\bf \S 8. Second Reconstruction Theorem}

\bigpagebreak

{\bf 8.1. Definition.} An abstract tree level system of correlation
functions (ACF) over a coefficient field $K$ consists of a pair
$(A,\Delta )$ as in Def. 6.1 and a family of even linear maps
 $$
Y_n:\ A^{\otimes n}\to K,\ n\ge 3 \eqno (8.1)
 $$
satisfying the following axioms:

\smallskip

{\bf 8.1.1. $S_n$--invariance.}

\smallskip

{\bf 8.1.2. Coherence.} In notation of 6.1 and (3.3) it reads:
for any pairwise distinct $1\le i,j,k,l\le n,$
 $$
\sum_{\{ ijSkl\}}\sum_{a,b} \varepsilon (S)
Y_{|S_1|+1}(\otimes_{r\in S_1}\gamma_r\otimes\Delta_a)g^{ab}
Y_{|S_2|+1}(\Delta_b\otimes (\otimes _{s\in S_2}\gamma_s))=
 $$
 $$
\sum_{\{ ikTjl\}}\sum_{a,b} \varepsilon (T)
Y_{|T_1|+1}(\otimes_{r\in T_1}\gamma_r\otimes\Delta_a)g^{ab}
Y_{|T_2|+1}(\Delta_b\otimes (\otimes_{s\in T_2}\gamma_s)).
\eqno (8.2)
 $$

\medskip

{\bf 8.2. Example.} For a tree level CohFT $I=(A,\Delta ,I_{0,n}),$
put
 $$
\langle I_{0,n}\rangle (\gamma_1\otimes\cdots\otimes\gamma_n):=
\int_{\overline{M}_{0,n}}I_{0,n}(\gamma_1\otimes\cdots\otimes\gamma_n).
\eqno (8.3)
 $$
These polynomials are called correlation functions of $I.$
Their $S_n$--symmetry is obvious, and (8.2) follows from the
Splitting Axiom 6.1 in the same way as (3.3) in the context
of GW--classes. The following main result of this section
shows that these examples essentially exhaust ACFs.

\medskip

\proclaim{\quad 8.3. Theorem} Any tree level ACF consists
of correlation functions of a unique tree level CohFT.
\endproclaim

\medskip

{\bf 8.3.1. Remark.} Starting with any ACF, one can construct
a potential
 $$
\Phi (\gamma )=\sum_{n\ge 3}\frac{1}{n!}Y_n(\gamma^{\otimes n})
 $$
and check that the differential equations (4.13) are equivalent to the
coherence relations (8.2). In this sense, WDVV equations
are equivalent to tree level CohFTs.

\smallskip

We start a proof of 8.3 with some preliminaries.

\medskip

{\bf 8.4. Correlation functions on trees.} In this context,
we will be considering tensors
 $$
\Cal{B}(\tau ):= A^{\otimes T_{\tau}}
\eqno (8.4)
 $$
rather than $\Cal{A}(\tau )=A^{\otimes F_{\tau}}.$ This construction
is obviously functorial with respect to pure contractions.
More important is its behaviour with respect to glueing and cutting.

\smallskip

If $(\tau ,e)=(\tau_1,t_1)*(\tau_2,t_2),$ we say that $(\tau_i, t_i)$
are obtained by cutting $\tau$ across $e.$ One can easily generalize
this notion for any subset of edges $E\subset E_{\tau}$ instead of
$\{ e\}.$ For instance, cutting $\tau$ across all edges results
in a set of one--vertex trees, {\it stars} of vertices $v\in V_{\tau}.$
Formally, a star $\rho (v,\tau )$ has $v$ as its vertex and
$F_{\tau}(v)$ as its tails.

\smallskip

Let $\tau |E$ be the set of trees obtained from $\tau$ by cutting
it across all $e\in E.$ Tails of any $\sigma \in \tau |E$
consist of some tails of $\tau $ and some ``half--edges''
of $\tau ,$ each edge in $E$ giving rise to two tails of the
latter type. Therefore, we can construct a well defined map
 $$
\Cal{B}(E):\ \Cal{B}(\tau )=A^{\otimes T_{\tau}}\to
\otimes_{\sigma\in\tau |E}\Cal{B}(\sigma )=
A^{\otimes (\coprod T_{\sigma})}\cong
A^{\otimes T_{\tau}}\otimes (A\otimes A)^E
\eqno (8.5)
 $$
which tensor multiplies any element of $\Cal{B}(\tau )$
by $\Delta^{\otimes E}\in (A\otimes A)^E.$
(Compare this to (6.6)).

\smallskip

\proclaim{\quad 8.4.1. Lemma} For any system of $S_n$--symmetric
polynomials $Y_n:\ A^{\otimes n}\to K,$ there exists a unique
extension to trees
 $$
Y(\tau ):\ \Cal{B}(\tau )\to K
 $$
with the following properties:

\smallskip

a). If $\rho_n$ is one--vertex tree with tails $\{ 1,\dots ,n\} ,$
then $Y(\rho_n)=Y_n.$

\smallskip

b). For any $\tau$ and any $E\subset E_{\tau},$ we have
 $$
Y(\tau )=\left( \otimes_{\sigma\in\tau |E}Y(\sigma )\right)\circ
\Cal{B}(E).
\eqno (8.6)
 $$

\smallskip

c). $Y(\tau )$ are compatible with tree isomorphisms.
\endproclaim

\smallskip

{\bf Proof.} Put
 $$
Y(\tau ):=\otimes_{\sigma\in\tau |E_{\tau}}Y(\sigma )\circ\Cal{B}(E_{\tau})=
\otimes_{v\in V_{\tau}}Y(\rho (v,\tau ))\circ \Cal{B}(E_{\tau}),
\eqno (8.7)
 $$
and $Y(\rho (v,\tau ))=Y_{|F_{\tau}(v)|}.$ Then a) follows by definition
from the $S_n$--symmetry of $Y_n,$ and (8.6) becomes a corollary
of the associativity of tensor products.

\smallskip

As an example of (8.6), let $\tau =\tau_S$ be a tree with vertices $v_1,v_2$
and tails $S_i$ ending at $i$--th vertex. Put $Y_{(i)}=Y(\rho (v_i,\tau )).$
Cutting
$\tau$ across its edge we get
 $$
Y(\tau )(\otimes_{i\in S}\gamma_i)=
(Y_{(1)}\otimes Y_{(2)})(\otimes_{r\in S_1}\gamma_r
\otimes\Delta\otimes (\otimes_{s\in S_2}
\gamma_s))=
 $$
 $$
\sum_{a,b}Y_{(1)}(\otimes_{r\in S_1}\gamma_r\otimes \Delta_a)
g^{ab}Y_{(2)}(\Delta_b\otimes (\otimes_{s\in S_2}\gamma_s)).
\eqno (8.8)
 $$

We turn now to correlation functions.

\medskip

\proclaim{\quad 8.5. Lemma} Let $\{ Y_n\}$ be correlation functions of
a CohFT I, $\{ Y(\tau )\}$ their extension to trees which we will call
operadic correlation functions.

\smallskip

For a tree $\tau$, denote by $f:\ \tau\to\rho$ a maximal pure contraction
identical on tails, and by $\varphi :\Cal{M}(\tau )\to\Cal{M}(\rho )$
the corresponding embedding. Then
 $$
Y(\tau )=\int_{\Cal{M}(\tau )}\varphi^*(I(\rho )).
\eqno (8.9)
 $$
\endproclaim
\smallskip

{\bf Proof.} From (6.4), we know that
 $$
\Cal{M}(\tau )=\prod_{v\in V_{\tau }}\Cal{M}(\rho (v,\tau )).
 $$
Write the relation (6.8) for $f$ taking in account the following
identifications:
 $$
\Cal{B}(\tau )=\Cal{A}(\rho )=A^{\otimes T_{\tau}},\
\Cal{A}(f)=\Cal{B}(E_{\tau}),\
 $$
 $$
H^*\Cal{M}(\tau )=\otimes_{v\in V_{\tau}}H^*\Cal{M}(\rho (v,\tau )),\
H^*\Cal{M}(f)=\varphi^*.
 $$
Thus, applying in addition (8.7),
we see that the following two functions $A^{\otimes T_{\tau}}\to K$
coincide:
 $$
\int_{\Cal{M}(\tau )}I(\tau )\circ \Cal{A}(f)=
\left(\otimes_{v\in V_{\tau}}\int_{\Cal{M}(\rho (v,\tau ))}
I(\rho (v,\tau ))\right)\circ \Cal{B}(E_{\tau})=Y(\tau )
 $$
and
 $$
\int_{\Cal{M}(\tau )}H^*\Cal{M}(f)\circ I(\rho )=
\int_{\Cal{M}(\tau )}\varphi^*(I(\rho )).
 $$

\medskip

\proclaim{\quad 8.5.1. Corollary} Let $f_{\alpha}:\ \tau_{\alpha}\to
\rho$ be a family of maximal pure contractions defining strata
$\Cal{M}_\alpha$ in $\Cal{M}(\rho )\cong \overline{M}_{0,n}$
whose homology classes $d_{\alpha}$ satisfy a linear
relation $R:\ \sum_{\alpha}a_{\alpha}d_{\alpha}=0.$

If $Y(\tau )$ are operadic correlation functions of a CohFT $I$,
they satisfy the identity
 $$
\sum_{\alpha}a_{\alpha}Y(\tau_{\alpha})=0.
\eqno (8.10)
 $$
\endproclaim

\smallskip

We will say that (8.10) is {\it correlated} with $R.$

\smallskip

For example, the Coherence Axiom 8.1.2 together with (8.6) means that
ACF satisfy all equations correlated with Keel's linear relations
between boundary divisors.

\smallskip

The central observation is that this implies the following
stronger statement:

\medskip

\proclaim{\quad 8.6. Lemma} Any ACF satisfies all the equations correlated with
linear relations between strata homology classes.
\endproclaim

\smallskip

{\bf Proof.} Clearly, it suffices to treat relations
$(7.1)_R$ where $R=(\tau ,\{ i,j,k,l\} ,v)$ as in (7.2). Consider
the star $\rho =\rho (v,\tau)$ and its four tails
$\overline{i},\overline{j},\overline{k},\overline{l}$ corresponding to
$e_i,e_j,e_k,e_l.$ Write the Keel relation in $H_*(\Cal{M}(\rho (v,\tau ))$
and the correlated identity (8.10):
 $$
\sum_{\{\overline{i}\overline{j}\rho^{\prime}\overline{k}\overline{l}\}  }
Y(\rho^{\prime})=
\sum_{\{\overline{i}\overline{k}\rho^{\prime\prime}\overline{j}\overline{l}\}}
Y(\rho^{\prime\prime})
\eqno (8.11)
 $$
where the sums are taken over two--vertex trees with tails $F_{\rho}.$

\smallskip

There is a natural bijection between summands in (8.11) and $(7.1)_R.$

\smallskip

Denote by $E$ the set of all interior edges of $\tau$ incident to $v$
excepting $e_i,e_j,e_k,e_l.$ Cut $\tau$ across these edges, and denote
by $T$ the set of resulting trees excepting the star of $v.$
According to (8.6), for the terms of l.h.s in (8.11) and $(7.1)_R$
corresponding to each other, we have
 $$
Y(\tau^{\prime})=Y(\rho^{\prime})\otimes\left( \otimes_{\sigma\in T}\right)
Y(\sigma )\circ \Cal{B}(E),
 $$
and similarly for the r.h.s.

\smallskip

Hence from (8.11) it follows that
 $$
\sum_{\{ ij\tau^{\prime}kl\}}Y(\tau^{\prime})=
\sum_{\{ ik\tau^{\prime\prime}jl\}}Y(\tau^{\prime\prime}).
 $$
This identity is correlated with $(7.1)_R.$

\medskip

{\bf 8.7. Proof of the Theorem 8.3.} Start with an ACF $Y_n.$ It
suffices to reconstruct an ``economy class'' CohFT defined by
6.1 rather than the full fledged operadic one. Therefore, in this section,
as in 7.1,
we may and will consider only $n$--trees and $n$--contractions.

\smallskip

First, construct $I(\rho_n)=I_{0,n}.$ From (8.9) we know the integrals
of $I(\rho_n)$ over all tree strata, whose homology classes generate
$H_*(\Cal{M}(\rho_n)),$ and by Lemma 8.6, these integrals
extend to a linear functional on $H_*(\Cal{M}(\rho_n)).$
By Poincar\'e duality, this uniquely defines $I_{0,n}.$

\smallskip

Second, check (6.1). It suffices to verify that integrals of
both sides over any stratum $\Cal{M}(\tau )$ contained in the
relevant boundary divisor $\Cal{M}(\tau_S)$ coincide.
But this is a particular case of (8.6). In fact, if $\tau =\tau_S,$
this is exactly (8.8). Generally,
one can assume that $(\tau ,e)=(\tau_1,t_1)*(\tau_2,t_2),\
T_{\tau_1}=S_1\cup \{ t_1\},\ T_{\tau_2}=S_2\cup \{ t_2\},$ and
apply (8.6) to this glueing.

\medskip

In the context of GW--classes, we can now more or less formally
deduce the following version of the Reconstruction Theorem 8.3.

\medskip

\proclaim{\quad 8.8. Theorem} Assume that for a manifold $V,$
a system of maps
 $$
Y^V_{n,\beta}:\ H^*(V,\bold{Q})^{\otimes n}\to \bold{Q}
 $$
is given satisfying all the conditions that are imposed
by Axioms 2.2.0--2.2.6 on the tree level codimension zero
GW-classes.

\smallskip

Then there exists a unique tree level system of GW--classes
$I^V_{0,n,\beta}$ such that $\langle I^V_{0,n,\beta}\rangle =
Y^V_{n,\beta }.$
\endproclaim

\medskip

To prove it, we first reconstruct the relevant CohFT, and then
check the Axioms involving the geometry of $V$. We leave the details
to an interested reader.

\smallskip

Thus starting with $V=\bold{P}^1$ (see (5.3) and (5.4)),
and applying the tensor product construction, we can
define GW--classes for $\bold{P}^1\times\dots\times\bold{P}^1.$
In particular, identities in 5.2.4 are thereby established.
Since $((\bold{P}^1)^n)^{S_n}\cong\bold{P}^n,$ we get
GW--classes for $\bold{P}^n$ as well.

\smallskip

Finally, results announced in [RT] give codimension zero
tree level GW-classes and therefore
all tree
level GW-classes for semi--positive symplectic manifolds.

\newpage

\centerline{\bf References}

\bigskip

[AM] P. S. Aspinwall, D. R. Morrison. Topological field theory
and rational curves. Preprint OUTP-91-32P, 1991.

\smallskip

[BCOV] M. Bershadsky, S. Cecotti, H. Ooguri, C. Vafa. Kodaira--Spencer
theory of gravity and exact results for quantum string amplitudes.
Preprint HUTP-93/A025.

\smallskip

[BG] A. Beilinson, V. Ginzburg. Infinitesimal structure of
moduli spaces of $G$--bundles. Int. Math. Res. Notices,
Duke Math. Journ., 4 (1992), 63--74.

\smallskip

[D] B. Dubrovin. Integrable systems in topological field theory.
Nucl. Phys. B 379 (1992), 627--689.

\smallskip

[GK] E. Getzler, M. M. Kapranov. Cyclic operads and cyclic homology.
Preprint, 1994.

\smallskip

[GiK] V. A. Ginzburg, M. M. Kapranov. Koszul duality for operads.
Preprint, 1993.

\smallskip

[I] C. Itzykson. Counting rational curves on rational surfaces.
Preprint Saclay T94/001.

\smallskip

[Ke] S. Keel. Intersection theory of moduli spaces of stable
$n$--pointed curves of genus zero. Trans. AMS, 330 (1992), 545--574.

\smallskip

[Ko] M. Kontsevich. $A_\infty$--algebras in mirror symmetry.
Bonn MPI Arbeitstagung talk, 1993.

\smallskip

[Ma1] Yu. Manin. Cubic forms: algebra, geometry, arithmetic.
North Holland, 1974.

\smallskip

[Ma2] Yu. Manin. Problems on rational points and rational curves
on algebraic varieties. (To be published in Surveys of Diff. Geometry).

\smallskip

[R] Y. Ruan. Topological sigma model and Donaldson type invariants
in Gromov theory. Preprint MPI, 1992.

\smallskip

[RT] Y. Ruan, G. Tian. Mathematical theory of quantum cohomology.
Preprint, 1993.

\smallskip

[V] A. Voronov. Topological field theories, string backgrounds, and
homotopy algebras. Preprint, 1993.

\smallskip

[W] E. Witten, Two-dimensional gravity and intersection theory
on moduli space. Surveys in Diff. Geom., 1 (1991), 243--310.

\smallskip

[Y] S.--T. Yau, ed. Essays on Mirror Manifolds. International Press
Co., Hong Cong, 1992.

\bigskip

{\it E--mail addresses:} maxim\@ mpim--bonn.mpg.de, manin\@ mpim--bonn.mpg.de

\enddocument